\documentclass[preprint,5p,twocolumn,10pt]{elsarticle}
\usepackage{amsmath,amssymb,amsfonts}
\usepackage{algorithmic}
\usepackage{graphicx}
\usepackage{textcomp}
\usepackage{balance}
\usepackage{xcolor}
\usepackage{booktabs}
\usepackage{amssymb}
\usepackage{enumitem}

\usepackage{enumitem}

\usepackage{multirow, makecell}

\usepackage{array, makecell}

\journal{Journal of Network and Computer Applications}
\begin{document}

\begin{frontmatter}

\title{On Challenges of Sixth-Generation (6G) Wireless Networks: A Comprehensive Survey of Requirements, Applications, and Security Issues}

\author[inst2]{Muhammad Sajjad Akbar}
\ead{muhammad.akbar@sydney.edu.au}
\author[inst1]{Zawar Hussain}
\ead{zawar.hussain@mq.edu.au}
\author[inst1]{Muhammad Ikram}
\ead{muhammad.ikram@mq.edu.au}
\author[inst1]{Quan Z. Sheng} 
\ead{michael.sheng@mq.edu.au}          
\author[inst1]{Subhas Mukhopadhyay}
\ead{subhas.mukhopadhyay@mq.edu.au}
\affiliation[inst2]{organization={School of Computer Science, Faculty of Engineering},
            addressline={University of Sydney}, 
            city={Sydney},
            postcode={2006}, 
            state={New South Wales},
            country={Australia}}
\affiliation[inst1]{organization={School of Computing, Faculty of Science and Engineering},
            addressline={Macquarie University}, 
            city={Sydney},
            postcode={2109}, 
            state={New South Wales},
            country={Australia}}            

\begin{abstract}
Fifth-generation (5G) wireless networks will likely offer high data rates, increased reliability, and low delay for mobile, personal, and local area networks. Along with the rapid growth of smart wireless sensing and communication technologies, data traffic has increased significantly and existing 5G networks are not able to fully support future massive data traffic for services, storage, and processing. To meet the challenges ahead, research communities and industry are exploring the sixth generation (6G) Terahertz-based wireless network that is expected to be offered to industrial users in just ten years.  
Gaining knowledge and understanding of the different challenges and facets of 6G is crucial in meeting future communication requirements and addressing evolving quality of service (QoS) demands. 
This survey comprehensively examines specifications, requirements, applications, and enabling technologies related to 6G. It covers disruptive and innovative, integration of 6G with advanced architectures and networks such as software-defined networks (SDN), network functions virtualization (NFV), Cloud/Fog computing, and Artificial Intelligence (AI) oriented technologies. The survey also addresses privacy and security concerns and provides potential futuristic use cases such as virtual reality, smart healthcare, and Industry 5.0. Furthermore, it identifies the current challenges and outlines future research directions to facilitate the deployment of 6G networks.
\end{abstract}


\end{frontmatter}


\section{Introduction} \label{sec:1}

The emergence of sixth-generation (6G) networks marks a pivotal moment in the evolution of wireless communication, poised to transcend the capabilities of its predecessor, 5G. As the torchbearer of the next generation of telecommunications, 6G promises to usher in a new era defined by higher frequencies, a substantial increase in capacity, and ultra-low latency~\cite{de2021survey, zong20196g}. Notably, 6G aspires to achieve an extraordinary milestone: one-microsecond transmission latency, a monumental advancement, surpassing the one-millisecond latency threshold of 5G networks by a factor of a thousand~\cite{markit2017internet}.

In addition to these impressive technical feats, 6G networks are poised to revolutionize a multitude of technological domains, including location awareness, imaging, and telepresence. A hallmark of 6G lies in its integration of machine learning and artificial intelligence (AI) technologies, which will pave the way for the development of autonomous systems within the 6G ecosystem. Leveraging the Terahertz (THz) spectrum, 6G sets its sights on achieving transmission rates of up to 1 terabyte per second (Tbps), offering an unprecedented network capacity and minimizing latency~\cite{zhang2018towards}.

\begin{figure}[htb!]
        \centering
            \includegraphics[width=0.48 \textwidth]{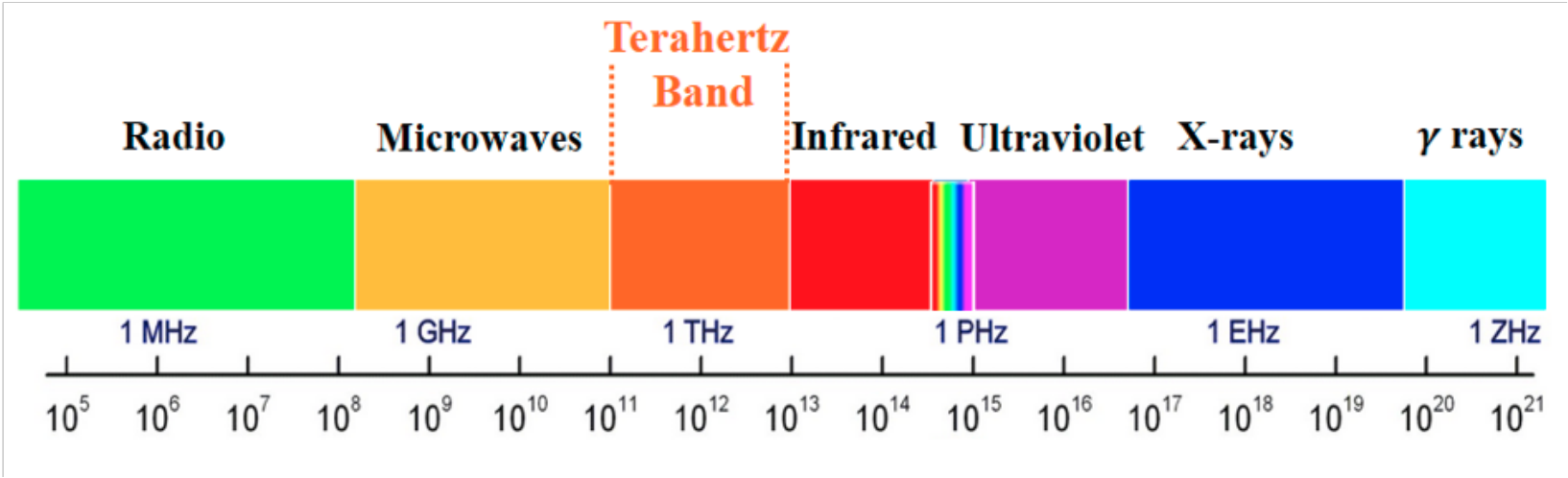}
            \caption{An overview of the Terahertz (THz) spectrum, ranging from 100 GHz to 3 THz, which is set to be utilized in 6G networks, adopted from~\cite{zhang2018towards}.}
            \label{Figure: THZZ}
\end{figure}

The advent of 6G networks not only promises exceptional network capacity but also enhances the transmission quality of existing applications. Using high frequencies, depicted in Figure~\ref{Figure: THZZ}, 6G networks enable increased sampling rates, thereby increasing the quality of wireless device transmissions.

The commercial availability of 6G networks is anticipated to become a reality in 2030, and substantial research and industrial efforts are already underway in this 
field.
Initiatives like the International Mobile Telecommunications 2030 project~\cite{david20186g} and China's Broadband Communications and New Networks project~\cite{khiadani2020vision} are shaping the trajectory of 6G networks. In the European Commission Horizon 2020 program, projects such as TERRANOVA~\cite{castro2022terranova} are exploring Tb/s wireless connectivity using the THz band, while research efforts in the United States, supported by the Semiconductor Research Corporation, investigate future communication over the THz band in cellular architecture~\cite{zong20196g}. Regulatory bodies like the Federal Communications Commission (FCC) are also actively investigating the potential of 6G networks utilizing the THz band. Furthermore, the International Telecommunication Union (ITU) has established a dedicated research group to define network requirements for the year 2030. 
In 2019, Finland took a proactive step by funding the 6Genesis project \cite{dardariinitial}, emphasizing the importance of looking beyond 5G to meet future quality of service (QoS) requirements \cite{series2015imt}. Each of these research endeavors highlights the importance of exploring and advancing beyond the capabilities of 5G to meet the evolving needs of future communication systems. 

{Unlicensed spectrum access is essential to the advancement of wireless communications and includes both the conventional sub-6 GHz and mmWave bands. The sub-6 GHz range is frequently used due to its advantageous propagation characteristics, which enable communications to travel farther and through obstructions more successfully. Wi-Fi (2.4 GHz and 5 GHz) and other wireless technologies like Bluetooth generally use this frequency. One such example is the 3.5 GHz Citizens Broadband Radio Service (CBRS), which balances licensed and unlicensed access to increase spectrum availability and efficiency by allowing shared use ~\cite{david20186g}.}

Compared to sub-6 GHz, the mmWave band (30-300 GHz) has a substantially larger bandwidth, enabling very high data rates required for cutting-edge applications like virtual reality (VR), augmented reality (AR), and streaming ultra-high definition video. However mmWave signals have limitations due to higher route loss and sensitivity to obstructions, which reduces their range and necessitates the use of technologies like massive MIMO and beamforming to improve performance. The use of unlicensed spectrum in these two bands is essential to increase the overall capacity of the network and network densification ~\cite{zong20196g}. To effectively manage these spectrum resources and provide reliable and scalable wireless communication networks, technologies like cognitive radio and dynamic spectrum sharing are crucial.

Unlike the deployment of 5G networks which faced several challenges, the forthcoming deployment of 6G networks will also face its own set of obstacles. Among these challenges, the development of commercial transceivers that operate on THz frequencies is a prominent concern. To address this challenge, it is crucial that electronics and networking component providers actively contribute by introducing new and innovative transceiver designs capable of operating in the THz range. 

Looking ahead, the proliferation of sensor-based systems is expected to witness the deployment of trillions of sensors in various settings such as cities, homes, and industries. This widespread adoption of sensor technology will give rise to a multitude of new applications. In light of this, the design of smart transceivers assumes a vital role in effectively managing the immense scale and complexity of these sensor-based implementations.

Considering these factors, it is evident that addressing the aforementioned challenges will play a crucial role in the successful deployment and management of 6G networks, especially in accommodating the extensive use of sensor-based systems in diverse domains. As 6G networks are still in the early stages of research and development, there is a need to systematically compile the existing knowledge and identify gaps in understanding. Moreover, given the limited existing literature that comprehensively addresses technical prerequisites, emerging applications, and critical security concerns for researchers, industry experts, and policymakers, this article embarks on an extensive exploration of these intricate dimensions associated with 6G networks.
This survey paper provides an up-to-date snapshot of the challenges and opportunities in 6G, serving as a valuable resource for researchers to pinpoint domains requiring deeper research and exploration.

\begin{figure*}[htb!]
        \centering
            \includegraphics[width=1\textwidth]{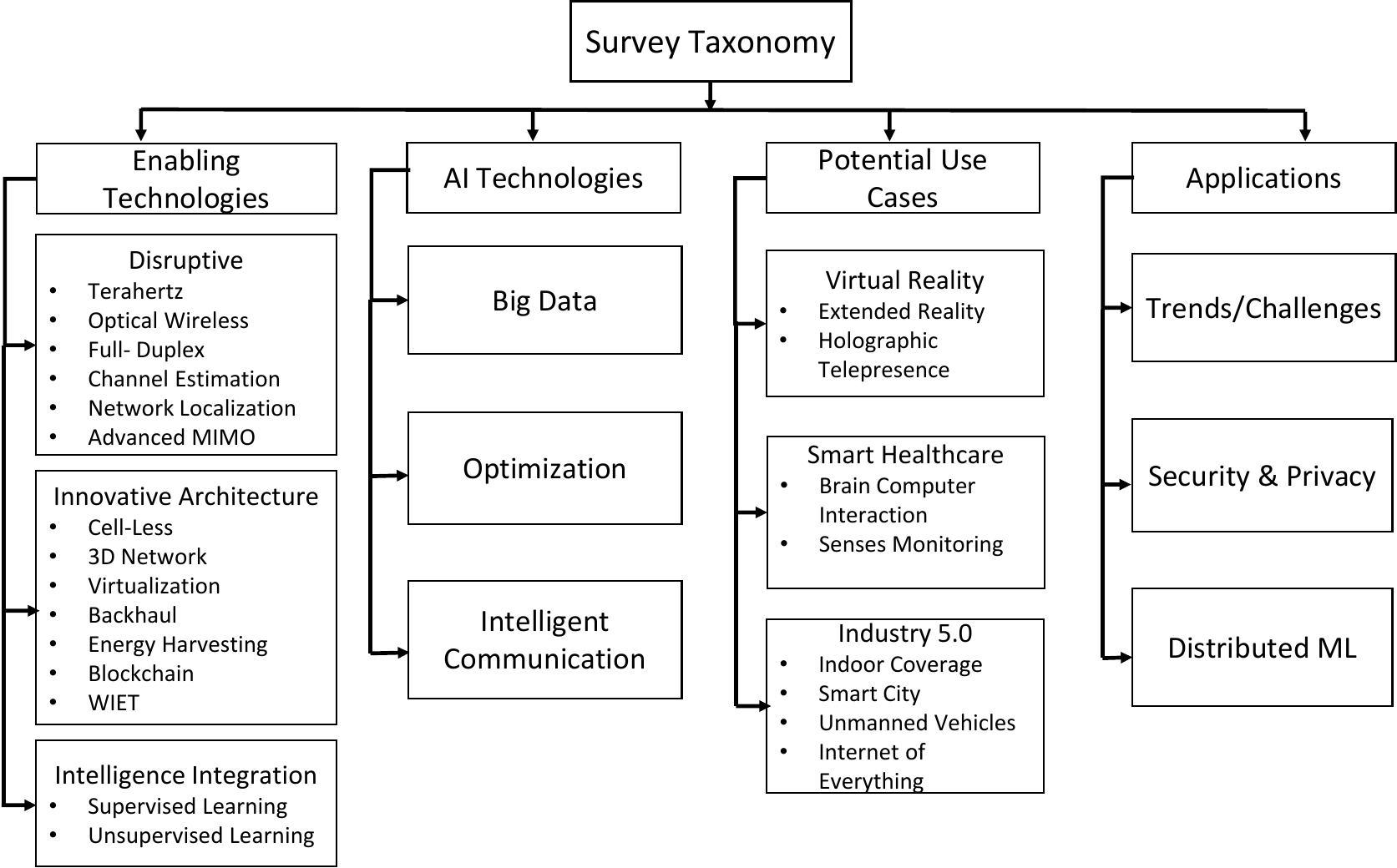}
            \caption{An overview of the taxonomy employed in this paper. In our endeavor to address the challenges of 6G wireless networks and conduct a comprehensive survey encompassing requirements, applications, and security issues, we classify the literature into four primary categories: enabling technologies, artificial technologies, case studies, and applications. We also provide detailed sub-classifications within these main categories. 
            } 
            \vspace{-0.5cm}
            \label{Figure:taxo}
\end{figure*}

{\bf Methodology.} This survey provides a comprehensive review of the various aspects of the 6G technology and the challenges associated with its adoption as documented in the existing literature. To ensure a thorough exploration of the subject, we begin by conducting an extensive search across major databases including ACM Digital Library, Science Direct, IEEE Xplore, and Scopus. Our search encompassed a wide range of topics related to 6G, including requirements, use cases, advanced architectures, integration, software-defined networking (SDN), Internet of Things (IoT), big data, machine learning (ML), and security issues, among others. 
For these sources, we then meticulously selected relevant papers from esteemed journals or venues such as IEEE Communications Surveys \& Tutorials, 
IEEE Vehicular Technology Magazine, IEEE Transactions on Wireless Communications, and IEEE Sensors. 

To ensure that our review encompasses the most current insights and advancements in the field of 6G, our investigation focused on relevant recent advancements. Therefore, we limited our selection to papers published within the last decade that specifically addressed various aspects of 6G. By adopting this approach, we aim to capture the latest developments and provide an up-to-date analysis of the field.


Next, we conducted a thorough examination of the titles and abstracts of the selected papers, categorizing the various aspects of 6G into four major classes. Our survey encompasses four key areas crucial to 6G: potential use cases, enabling technologies, AI technologies, and applications. The taxonomy of this survey is visually represented in Figure \ref{Figure:taxo}. The discussion on potential use cases further delves into three distinct aspects: virtual reality, smart healthcare, and Industry 5.0. Enabling technologies are subdivided into three main categories: innovative architectures, disruptive technologies, and integration of intelligence. AI technologies are explored in terms of big data, optimization of existing algorithms, and examples of intelligent communication. We also address various 6G-based applications with challenges, security, and privacy considerations, and distributed machine learning. 

By reviewing the state-of-the-art works, we aim to comprehensively cover all these aspects of 6G networks, providing users with a comprehensive overview of the opportunities, challenges, and requirements associated with the 6G network. Table~\ref{tab: PrevSurveys} provides a brief comparison of this survey paper with existing 6G survey papers. Additionally, to enhance readability, we have compiled a list of acronyms used throughout this paper in Table \ref{tab:acronyms}.

\subsection{Contributions}

This survey 
tracks the path towards 6G comprehensively by considering a broad range of aspects mentioned in Figure \ref{Figure:taxo} and the main contributions are as follows:

\begin{itemize}
	\itemsep -1pt

\item {\textbf{Comprehensive comparative analysis.} Most of the previous surveys~\cite{shahraki2021comprehensive, chowdhury20206g, yang20196g, alsharif2020sixth, akyildiz20206g, zhang20196g, you2021towards, khan20206g, zhao2021survey, khanh2023innovative, serghiou2022terahertz, al2023edge} used only ``Yes'' or ``No'' for comparing different aspects of the 6G ecosystem which hardly provides any details. Unlike the previous surveys, this work provides (Section ~\ref{sec:2}) a fine-grained comparison like Low (L), Medium (M), High (H), and No (N) for different factors. In essence, we correlate the various technologies and architectures (such as Blockchain, Industry 5.0, augmented reality (AR), mixed reality (MR), virtual reality (VR), mmwave, Unmanned Aerial Vehicle (UAV), software-defined networks (SDN), and many more) with 6G networks by addressing their integration issues such as required data rate in terabyte per second (Tbps), minimum specified delay, mobility, energy efficiency, and spectrum issues.}

\item {\textbf{Outline the requirements and vision for 6G.} 
By 2030, a new generation of mobile communication networks is expected, driven by sociological and technological developments discussed in this article. The discourse elucidates the emergence of various novel applications and details how the capabilities of forthcoming 6G networks will facilitate their realization. Taking into account the challenges of building 6G and the lessons learned during the development of 5G (Section ~\ref{sec:3}), we offer a roadmap for future 6G research directions. Our paper outlines the anticipated requirements for 6G (Section \ref{sec:3}) and conducts a comprehensive discussion on the enabling technologies that will meet these requirements for upcoming applications (Section ~\ref{sec:4}).}

\item {\textbf{Use cases categorization.} Our survey offers an extensive classification of diverse 6G-related usecases (Section \ref{sec:6}). Specifically, these scenarios are grouped into categories such as VR, Healthcare, Industry 5.0, as well as enabling technologies, both disruptive and innovative, and AI, encompassing machine learning and big data.}
 \end{itemize}

{\bf Organization.} The remainder of our paper is structured as follows: Section~\ref{sec:2} discusses the related work and Section ~\ref{sec:3} provides the background for communication technologies. Section ~\ref{sec:4} describes the 6G key enabling technologies while in Section~\ref{sec:5}, we provide a discussion on AI-enabled technologies and applications for future 6G. Section~\ref{sec:6} presents the potential use cases of 6G technology. 
Section~\ref{sec:8} discusses the challenges in the 6G network and the future research directions, and Section~\ref{sec:9} concludes our work.

\begin{table}[!th]
\caption{\label{tab:acronyms} List of acronyms and the corresponding definitions.}
\scalebox{0.92} {
\begin{tabular}{p{1.3cm}|p{2.6cm}|p {1.0cm}|p{2.4cm}} \hline
\textbf{Acron.} & \textbf{Definition}                              & \textbf{Acron.} & \textbf{Definition}                       \\ 
\hline

1/2/3 /4/5/6 G      & First/Second /Third/Fourth /Fifth/Sixth Generation & Tbps              & Terabyte per second                       \\ 
\hline
AI                & Artificial Intelligence                          & THz               & Tera Hertz                                \\ 
\hline
3GPP              & 3rd Generation Partnership Project               & GHz               & Giga Hertz                                \\ 
\hline
mmWave            & millimeter-wave                                   & FCC               & Federal Communications Commission         \\ 
\hline
QoS               & Quality of Service                               & NMR               & Nordic Mobile Radio System                \\ 
\hline
GSM               & Global System for Mobile Communication           & SMS               & Short Messaging Service                   \\ 
\hline
EDGE              & Enhanced Data Rates for GSM Evolution            & GPRS              & General Packet Radio Service              \\ 
\hline
UMTS              & Universal Mobile Telecommunication System        & CDMA              & Code Division Multiple Access             \\ 
\hline
D2D               & Device to Device                                 & LTE               & Long-Term Evolution                       \\ 
\hline
ML                & Machine Learning                                 & VR                & Virtual Reality                           \\ 
\hline
MR                & Mixed Reality                                    & AR                & Augmented Reality                         \\ 
\hline
UAV               & Unmanned Aerial Vehicle                          & eMBB              & Enhanced Mobile Broadband                 \\ 
\hline
SDN               & Software Defined Network                         & URLLC             & Ultra-Reliable low Latency Communications  \\ 
\hline
FSO               & Free Space Optical                               & IoT               & Internet of Things                        \\ 
\hline
BCI               & Brain Computer Interaction                       & IoE               & Internet of Every Thing                   \\ 
\hline
HCI               & Human-Computer Interaction                       & M2M               & Machine to Machine                        \\ 
\hline
DAS               & Distributed Antenna Systems                      & UGV               & Unmanned Ground Vehicle                   \\ 
\hline
MIMO              & Multiple-Input-Multiple-Output                   & UUV               & Unmanned Underwater Vehicle               \\ 
\hline
VLC               & Visible Light Communication                      & RF                & Radio Frequency                           \\ 
\hline
NFV               & Network Functions Virtualization                 & WIET              & Wireless Information and Energy Transfer  \\
\hline
\end{tabular}
}
\end{table}

\section{Comparison with Existing Surveys} \label{sec:2}

Several surveys have reviewed the literature conducted in the area of 6G. These surveys have focused on different aspects of 6G technologies. This section provides an overview of some of the most recent surveys and also provides a comparison with our work.

The survey  \cite{shahraki2021comprehensive} discusses the 6G-based features and challenges for Cellular IoT which can connect enormous physical and virtual objects to the Internet using cellular networks. These issues include delay and throughput provision for bandwidth-hungry applications like holographic communication etc. The survey also discusses various aspects of 6G networks like needs, requirements, applications, key performance indicators, and the potential key enabling technologies. However, it does not provide details about security, ML/AI, and various potential use cases like Industry 5.0.

The authors in \cite{chowdhury20206g} discuss about the reasons behind 6G's development and the expected uses for its cutting-edge capabilities. It addresses several possible uses, such as the Internet of Things (IoT), autonomous systems, smart cities, virtual and augmented reality (AR, VR), and healthcare. A thorough list of applications is provided by the survey, however, it falls short of providing detailed analysis and concrete examples to back up the statements made regarding their viability and possible impact. Ultra-high data speeds, ultra-low latency, huge connection, energy efficiency, and improved security and privacy are just a few of the features that the authors list as necessary for 6G. Nevertheless, the study neglects to assess these requirements critically in light of potential technical advances and trade-offs that may be required to meet them. A more comprehensive analysis of the viability and obstacles related to fulfilling these demands would have enhanced the depth of the conversation. A few potential technologies that could be used in 6G systems are mentioned in passing in the article, including enhanced antenna systems, terahertz (THz) communications, millimeter-wave (mmWave) communications, and artificial intelligence (AI). However, the conversation about these technologies is still quite cursory and doesn't go into great detail about their strengths, weaknesses, and possible implementation issues. Although difficulties and future possibilities for research are mentioned in the paper, its coverage is somewhat narrow and does not provide a critical assessment of the major roadblocks that must be removed to successfully implement 6G networks.

The authors of \cite{yang20196g} highlight the need for ultra-high data speeds, ultra-low latency, huge connection, and dependable communication to support new technologies such as IoT, artificial intelligence (AI), and virtual reality (VR). Although the report offers a broad vision for 6G, it does not critically assess whether or not these lofty objectives can be achieved. The authors present some possible methods that might be used in wireless communications for 6G. These include massive multiple-input multiple-output (MIMO), non-orthogonal multiple access (NOMA), terahertz (THz) communications, and communication systems supported by artificial intelligence (AI). Nevertheless, the study simply offers a cursory summary of these methods; it doesn't explore their unique benefits, constraints, and implementation difficulties. A more perceptive viewpoint might have been possible with a more critical examination of the trade-offs and possible problems related to these strategies. The claims stated about the prospective techniques are not supported by empirical evidence or by any particular examples in the study. Case studies or simulations to show the efficiency and performance of these methods in actual situations would have been helpful. Furthermore, a detailed examination of the difficulties and the prospects of 6G wireless communications research is absent from the work. Although the writers touch on a few issues, like spectrum allocation, energy efficiency, and security, their coverage is somewhat scant and does not include a thorough analysis of the main problems and viable solutions.

The paper \cite{alsharif2020sixth} describes the goals of 6G, which include vast connection, ubiquitous coverage, ultra-high data speeds, ultra-low latency, and the seamless integration of several technologies. While the survey gives a broad vision, it lacks critical analysis and specific data to support the feasibility and practicality of accomplishing these aims. The writers talk about current studies being conducted in the area of 6G wireless networks. They highlight themes such as millimeter-wave (mmWave) communications, massive multiple-input multiple-output (MIMO), terahertz (THz) communications, energy-efficient designs, and artificial intelligence (AI)-enabled networks. Nevertheless, the paper's coverage of these research activities is still cursory, and it does not provide a thorough analysis of the advancements made, the shortcomings of the current solutions, or possible future research areas. The issues of 6G wireless networks, such as spectrum management, energy efficiency, security, and privacy, are briefly covered in this study. Nevertheless, there isn't a thorough examination of the main barriers that need to be removed, and the discussion of these difficulties is somewhat restricted. Furthermore, the reader is left desiring a more thorough investigation because the suggested remedies to these problems are not fully developed. A critical assessment of the possible effects, advantages, and disadvantages of 6G wireless networks is also absent from the report. It doesn't fully look at the risks and trade-offs connected to the lofty objectives and suggested fixes. A more balanced assessment of the merits and downsides of 6G would have provided a more comprehensive knowledge of the matter.

The paper \cite{akyildiz20206g} draws attention to the need for future wireless systems to have greater data rates, reduced latency, huge connectivity, and energy efficiency. Nevertheless, a critical evaluation of these demands and the potential trade-offs in attempting to meet them is absent from the work. The writers examine various methods and technologies that might be essential to the advancement of 6G and other technologies. Terahertz (THz) communications, software-defined networking (SDN), intelligent reflecting surfaces (IRS), and visible light communications (VLC) are a few of these. Although these technologies are briefly reviewed in the paper, their viability, difficulties in implementation, and potential drawbacks are not fully assessed. Most of the discussion is still theoretical, and the assertions made are not backed up by specific instances or data. Moreover, a detailed examination of the possible effects and societal ramifications of upcoming wireless communication technologies is absent from the report. It falls short of addressing the difficulties and worries about security, privacy, and other legal issues that could come up when new technology is implemented. Additionally, the article fails to highlight the research activities, active initiatives, and standardization efforts in the field of 6G and beyond. Giving an overview of the state of research and the advancements made by different stakeholders in the field of wireless communication technology would have been beneficial.

The authors \cite{zhang20196g} describe the vision for 6G, highlighting features such as ultra-low latency, ultra-high data speeds, extensive interconnectedness, and pervasive intelligence. While the vision is broad and ambitious, the article lacks critical analysis and concrete data to support the feasibility and practicality of achieving these high goals. The authors discuss the specifications for 6G wireless networks, including improved energy and spectral efficiencies, sustainability, reliability, and security. However, there is minimal in-depth examination of the trade-offs and challenges involved in meeting these requirements, with the discussion remaining at a high level.

A more thorough understanding would have been achieved by critically analyzing the requirements and their potential impacts on network performance and implementation. The study presents a high-level architecture for 6G wireless networks, incorporating edge computing, terahertz (THz) communications, and artificial intelligence (AI) among other heterogeneous technologies. However, the architecture is only conceptually explored, lacking detailed analysis of specific design considerations, implementation challenges, and protocols. A more comprehensive examination of the architectural components and their interactions would have enhanced the paper's value.

Additionally, the report covers key technologies such as blockchain, machine learning, cognitive radio, and advanced antenna systems that could be significant in 6G wireless networks. However, the article does not delve deeply into these technologies nor offer a critical assessment of their advantages, disadvantages, and applicability to 6G.

The authors \cite{you2021towards} highlight the necessity of 6G networks to meet the growing demands of cutting-edge applications, such as IoT, virtual reality, and ultra-high-definition video. While the paper presents a comprehensive vision, it lacks critical analysis and specific evidence to support the feasibility and practicality of achieving the defined goals. The authors discuss enabling technologies, including massive multiple-input multiple-output (MIMO), edge computing, terahertz (THz) communications, and artificial intelligence (AI), which could be crucial to the development of 6G networks. However, the discussion of these technologies remains at a high level and does not delve into their drawbacks, application challenges, and potential trade-offs. A more critical review of these technologies and their relevance to 6G networks would have enhanced the paper's contribution.

The study examines novel paradigm shifts in network intelligence, security and privacy, sustainability, and user-centric design that may influence the future of wireless communication. However, the research does not fully explore the possible implications, advantages, and disadvantages of these paradigm shifts, and the discussion is somewhat scant. A more thorough examination of the consequences and trade-offs associated with these changes would have provided a deeper understanding of their significance for 6G networks.

The paper \cite{khan20206g} explains advanced features such as intelligent networking, massive connectivity, ultra-low latency, and ultra-high data speeds. Although the article offers a bold vision, it lacks critical analysis and concrete data to support the feasibility and practicality of achieving these objectives. The authors discuss innovative waveform designs, edge computing, AI, and intelligent reflecting surfaces (IRS) as architectural components that may be essential in the creation of 6G systems. However, there is little consideration of these components' potential drawbacks, practical applications, and related challenges. A more rigorous assessment of these components and their interactions within the 6G architecture would have provided a more thorough understanding.

The article also explores future directions for 6G systems, including the incorporation of mmWave, blockchain, and satellite communication. However, the research does not fully explore the benefits, drawbacks, and viability of these strategies, and the discussion remains theoretical. A more critical assessment of the possible advantages, disadvantages, and implementation difficulties related to these future paths would have enhanced the paper's contribution. Furthermore, it falls short of addressing the challenges and concerns regarding security, privacy, and regulatory issues that could arise with the implementation of new technologies.

The authors \cite{zhao2021survey} highlight the need for ultra-high data speeds, ultra-low latency, massive connectivity, and intelligent networking, discussing the reasoning behind and future goals of 6G wireless communications. Although the vision is extensive, the article lacks critical analysis and concrete data to support the feasibility and practicality of achieving these lofty objectives. The authors provide a comprehensive overview of cutting-edge technologies such as THz communications, non-orthogonal multiple access (NOMA), visible light communications (VLC), AI, and quantum communications that may be relevant to 6G networks. However, the study does not thoroughly analyze or review the benefits, drawbacks, and potential challenges of these technologies. Instead, the discussion remains at a high level.

The article would have contributed more by providing a more rigorous evaluation of these technologies and their suitability for 6G networks. The application of these advanced technologies in 6G networks involves numerous open research questions and significant hurdles that are not fully explored in this paper. Issues such as security, energy efficiency, scalability, and interoperability are not sufficiently covered. A more critical evaluation of the potential roadblocks and knowledge gaps would have provided valuable insights for further studies.

The paper \cite{khanh2023innovative} discusses the need for 6G and underlines the possible advancements it may offer in terms of ultra-high data speeds, ultra-low latency, huge connection, and intelligent networking. Although the article offers an ambitious vision for 6G, it is devoid of critical analysis and hard data to back up the viability and usefulness of reaching these objectives. The writers offer an overview of the many technological advancements, applications, and architectural components that are anticipated to be significant in the 6G future. These include ideas like ambient intelligence, network slicing, holographic communications, and enhanced sensing methods. Nevertheless, these subjects are only briefly discussed, and the study does not provide a thorough examination or assessment of their applicability in real-world settings, any drawbacks, or related difficulties. The paper's contribution would have been strengthened by a more critical evaluation of these patterns and their implications for 6G networks.  Although the study mentions the difficulties in deploying 6G technology, it does not go into great detail to address these difficulties. It doesn't go into detail about how difficult it is to integrate and standardize new technology, how to handle privacy and security issues, or how 6G networks may affect society and the law. A greater comprehension of the potential roadblocks to the actualization of the 6G era would have been possible with a more critical analysis of these difficulties.

The authors \cite{serghiou2022terahertz} highlight the importance of THz frequencies in the context of 6G wireless communications, discussing their potential to provide exceptionally high data rates and bandwidths. While the study recognizes the potential benefits of THz communications, it does not offer a thorough analysis or hard data to support the viability and utility of using THz frequencies in 6G networks. The phenomena of THz channel propagation, including attenuation, path loss, and dispersion, are reviewed by the authors, who also discuss the concepts and measurement methods used to describe THz channels. However, there is no critical assessment of the models and measurement methods; instead, the publication primarily focuses on summarizing existing research.

A more comprehensive understanding of the challenges involved in THz channel characterization would have been possible with an in-depth examination and comparison of these methods, including an awareness of their limitations and uncertainties. Additionally, the study briefly discusses unresolved issues and potential avenues for future THz communications research for 6G applications. Although it touches on topics like beamforming, channel modeling, and antenna design, it does not delve deeply into the specific challenges and unmet research needs in these areas. The paper's contribution could have been enhanced by a more critical analysis of the unresolved issues and a thorough discussion of possible solutions.

The paper \cite{al2023edge} explains the growing interest in implementing intelligent network edge capabilities to meet the needs of future wireless communication systems. It discusses federated learning and edge-native intelligence as enabling technologies for efficient data processing and decision-making at the network edge. However, the study does not critically evaluate the shortcomings and potential disadvantages of these strategies. The paper's contribution could have been strengthened with a more thorough assessment of the difficulties and trade-offs related to edge-native intelligence and federated learning.

The authors provide an overview of the developments and trends in federated learning and edge-native intelligence, covering privacy-preserving techniques, optimization methods, and distributed learning. While the paper effectively summarizes these procedures, it does not critically analyze or compare various methods. A more detailed examination of the advantages and disadvantages of different approaches would have offered a deeper understanding of their suitability within the framework of 6G communications.

Additionally, the study briefly discusses the challenges and potential avenues for future research in this area. It touches on issues such as heterogeneous data, communication limitations, and privacy and security challenges. However, the study only scratches the surface of these issues, lacking detail on the specific practical and technological difficulties involved in addressing them. The paper's contribution would have been enhanced by a more thorough investigation of prospective solutions and a more critical assessment of the open research directions.

The paper \cite{qadir2022towards} explains the evolution of wireless networks from earlier generations to 6G, emphasizing the necessity of ultra-reliable and low-latency communication to meet the diverse needs of Internet of Things (IoT) applications. It highlights the potential of 6G to offer faster data speeds, better coverage, and improved energy efficiency. The authors explore the latest developments in the field, covering key supporting technologies such as network slicing, ultra-reliable and low-latency communications (URLLC), and massive machine-type communications (MTC). They discuss how these advancements can enable new applications in industrial automation, smart cities, healthcare, and transportation.

Though the paper provides a broad overview of current developments and applications, it lacks critical analysis and in-depth examination. It does not offer sufficient evidence or concrete examples to support the claims made about the features and benefits of 6G IoT. Additionally, the open challenges are only briefly mentioned, and alternative solutions are not thoroughly explored. The report also fails to discuss the potential downsides or limitations of 6G IoT, such as security, privacy, and ethical issues that arise as more IoT devices are integrated into various aspects of daily life. A more careful analysis of the possible trade-offs and risks associated with 6G IoT would have added depth and complexity to the discussion.

The paper \cite{zhou2023aerospace} demonstrates the aircraft industry's growing need for dependable, high-performing communication solutions. It examines the concept of aerospace-integrated networks and their potential to accommodate various services and applications in 6G networks. However, a critical evaluation of the viability and usefulness of deploying aerospace-integrated networks in the context of 6G is missing from the research. A more comprehensive assessment of the technical obstacles, financial considerations, and regulatory aspects involved with aerospace-integrated networks would have enhanced the paper's contribution.

The authors provide an overview of the major architectures and technologies used in aerospace-integrated networks, including unmanned aerial vehicles (UAVs), aerial platforms, and satellite communication systems. While the report effectively summarizes these technologies, it does not thoroughly analyze or assess their impact on 6G networks. A more rigorous evaluation of the advantages and drawbacks of aerospace-integrated networks, including their scalability, reliability, and compatibility with existing terrestrial networks, would have provided a deeper understanding of their potential in the 6G era.

The report also briefly mentions future research areas and challenges in the field, such as resource allocation, interference reduction, and spectrum management. However, it only scratches the surface of these issues and does not delve into the technical details and practical challenges involved in addressing them. A more critical examination of the open research directions and a detailed analysis of potential solutions would have strengthened the paper's contribution.

The paper \cite{mogyorosi2022positioning} emphasizes the importance of accurate positioning for future wireless communication systems and explores the potential uses and applications that depend on precise location information. However, the study does not critically examine the constraints and challenges related to positioning in 5G and 6G networks. A more thorough assessment of the precision, reliability, and scalability of current positioning methods would have enhanced the paper's value.

The authors provide an overview of different positioning technologies and methods, such as hybrid approaches, cellular-based techniques, and satellite-based systems, discussing network-based location, time-of-flight measurements, and global navigation satellite systems (GNSS). While the paper effectively summarizes these tactics, it does not critically assess or compare various approaches. A more detailed investigation of their advantages and disadvantages, including performance in different environments and susceptibility to interference and multipath effects, would have offered a deeper understanding of the suitability of different positioning technologies for 5G and 6G networks.

The study also briefly discusses the challenges and potential paths for 5G and 6G network positioning, addressing issues like indoor positioning, hybrid positioning methods, and multipath mitigation. However, the discussion is brief and lacks detail on the specific technical difficulties and practical issues involved. A more thorough investigation of prospective solutions and a more critical assessment of the open research directions would have enhanced the paper's contribution.

The authors \cite{wang2023road} discuss the goals and motivations behind the development of 6G networks, emphasizing the necessity of ultra-low latency, massive connectivity, sophisticated networking capabilities, and significantly higher data rates. However, the report lacks a critical evaluation of the feasibility and practicality of achieving these lofty objectives. To enhance the paper's value, a more thorough analysis of the technological obstacles, financial implications, and regulatory considerations related to the deployment of 6G networks would have been necessary.

The authors present the main specifications and performance goals for 6G networks, covering topics like intelligent networking, integrated sensing and communication, and terahertz (THz) communication. While the report does a decent job of outlining these requirements, it does not critically assess their feasibility or implications for network infrastructure. A more thorough examination of the trade-offs and challenges in meeting these requirements would have provided a deeper understanding of the difficulties encountered in developing 6G networks.

The study also addresses major supporting technologies for 6G networks, such as edge computing, massive multiple-input multiple-output (MIMO), THz communication, and AI. Although the study provides a thorough review of these technologies, it does not critically analyze their integration into existing network infrastructures, scalability, or practical implementation issues. A more comprehensive analysis of potential drawbacks, interoperability problems, and standardization efforts for these crucial technologies would have increased the paper's value.

Additionally, the study provides a brief overview of 6G experimental platforms and testbeds. However, it does not go into detail about the testbeds currently in use or offer a critical assessment of their strengths and weaknesses. A more thorough analysis of the opportunities and practical challenges associated with developing and deploying 6G testbeds would have benefited the report.

The paper \cite{fadlullah2022balancing} highlights the growing importance of edge computing and its potential to enable a range of novel applications, emphasizing the need for secure and reliable services in edge environments. However, it does not provide a thorough analysis of the specific challenges and trade-offs involved in achieving this balance. The paper's contribution could have been enhanced by a more detailed examination of the competing demands between security and quality of service (QoS), as well as their potential impacts on network performance and user experience.

The authors outline current methods and approaches for balancing edge computing security and QoS, discussing topics such as resource management, anomaly detection, and intrusion detection. While the study offers a decent summary of various approaches, it lacks critical analysis and comparison of their efficacy, scalability, and practicality. A more in-depth evaluation of the limitations, deployment issues, and potential trade-offs of these techniques would have provided valuable insights.

The study also explores how machine learning approaches can assist in managing the trade-offs between security and QoS in edge computing. It discusses the use of machine learning for dynamic resource allocation, anomaly detection, and intrusion detection. However, the study does not include case studies, empirical data, or specific examples to support the claim that machine learning is an effective tool for achieving this balance. A more thorough assessment of the effectiveness, resilience, and suitability of machine learning algorithms in real-world edge computing scenarios would have strengthened the paper's contribution.

The study emphasizes the potential of 6G networks to address edge computing's security and QoS challenges, proposing the need for more intelligent, self-sufficient, and context-aware edge systems to adapt dynamically to changing network conditions. While the study presents several intriguing concepts and ideas for 6G, it does not critically examine the feasibility and practicality of implementing these ideas. The article would have been more comprehensive if it had explored the operational and technological challenges, legal issues, and standardization efforts required to realize these 6G prospects.

The authors \cite{chafii2023twelve} enumerate various issues present in 6G networks, such as waveform design, artificial intelligence (AI) integration, channel modeling, interference management, energy efficiency, and security. They emphasize the need to reconsider conventional methods and develop novel approaches to overcome the limitations of current communication theories. However, while the study presents a long list of issues, it does not analyze or examine each one in great detail. A more thorough explanation of the difficulties and their potential impact on 6G networks, supported by real-world examples or case studies, would have been beneficial.

Furthermore, the document lacks a defined framework or roadmap, offering no guidance to the reader on how to approach and handle each of these challenges. The article briefly mentions the integration of AI approaches into 6G networks but does not delve into their specific applications or how they can address the identified problems. A more in-depth analysis of AI-driven solutions and their potential impact on resolving these issues would have greatly enhanced the research.

Additionally, the paper does not include implementation details or practical implications of the proposed solutions. Investigating the feasibility, scalability, and potential trade-offs of applying these strategies in actual 6G networks would have added significant value to the study.

The article \cite{guo2021enabling} provides a comprehensive analysis of the potential for massive IoT enabled by 6G. It thoroughly examines the requirements and driving forces behind new IoT-enabled applications, highlighting the limitations of both 5G and 6G. The authors also delve into key areas of 6G related to IoT, including technical specifications, application scenarios, and emerging trends. The authors propose an Internet of Things architecture known as space-air-ground-underwater/sea networks, which integrates machine learning techniques and is anticipated to leverage 6G technologies. However, the article overlooks security concerns specific to 6G and fails to address several potential use cases such as virtual reality, Industry 5.0, big data applications, and health applications in its survey.

Wang et al., \cite{wang20206g} gives performance evaluation metrics for 6G wireless communication networks and presents a perspective of the recent paradigm shift in these networks. The survey's main goal is to present a thorough examination of channel characteristics for mm-wave, terahertz, and optical wireless communication, among other 5G and 6G frequency channels. The satellite, UAV, marine, and undersea application scenarios are covered by these channel parameters. The paper made note of the issues that 6G channel measurements and models will face in the future, however, it lacked information on disruptive technologies based on 6G. A crucial component of 6G networks, the security issues on these channels were also disregarded in the essay.

The articles \cite{mondal2015survey, gawas2015overview} review the specifications, benefits, and drawbacks of the various generations of mobile wireless technology. The writers concentrated on channel modeling, particularly the 6G networks' channel modeling possibilities. Additionally, a cost-benefit analysis is given. Nevertheless, the integration of the channel models with 6G networks is not discussed in the publications.
 
Free space optical (FSO) \cite{esmail2016outdoor} is an emerging technology for 6G networks and has the potential to resolve the bandwidth bottleneck. The authors discussed the issues caused by the fog’s attenuation to FSO and they proposed an empirical prediction model to calculate its impact. 

The goal of the survey \cite{ vinogradov2016key} aims to present the current status of UAVs and comprehend the problems with telecoms that need to be fixed. The implementation suggested for UAVs is detailed together with a schematic architecture. The author analyzes Automatic Dependent Surveillance-Broadcast (ADS-B) to 5G/6G Cellular Networks with Special Reference to Aviation. The authors only covered the essential needs of 6G networks for unmanned aerial vehicles (UAVs); comprehensive use cases and security considerations need to be supplied. 

In \cite{ ankarali2017flexible}, a framework for creating adaptable radio access technologies (RATs) for 6G networks and the associated specifications was provided by the authors. The framework allows for freedom in RAT selection. The inefficiencies of the fixed waveform parameterization of 5G and 6G numerologies are further highlighted by this approach. RAT security concerns aren't discussed in the article, though.

The survey \cite{al2017proposed} gives an overview of the growth of wireless networks, including the requirements and problems of the 5G and 6G heterogonous network architecture. The integration of cutting-edge 6G technologies, such as Machine to Machine Communications (M2M) and Software-Defined Networking (SDN), in the framework of 5G and 6G networks was also covered by the writers. Nevertheless, a thorough explanation of how these domains are integrated with 6G networks is absent from the survey.

The authors in \cite{ reka2018future} present a complete survey on the Smart grid based on IoT frameworks. To assess such a framework, several performance indicators have been identified; these metrics include high throughput, power management, security, latency, etc. The authors concluded that not enough was being done to secure Internet of Things-based frameworks and recommended that cyber security be made a requirement for these systems going forward. There isn't a discussion of 6G applications, specifications, how to integrate IoTs with sophisticated network designs, etc. in the poll.
  
The survey \cite{ yastrebova2018future} provides a thorough analysis of the specifications and designs of 6G networks. To determine whether the suggested topologies were appropriate for 6G networks, the authors assessed them using performance metrics including throughput and latency. Many applications are used for the evaluation, including sensitivity (touch, smell, texture, etc.) and senses like 3D video. The authors concluded that Super Ultra-Low Latency, precision latency, and ultra-delays were necessary for 6G networks to support the applications listed.

The survey \cite{ long2019software} offers a thorough analysis of a cellular system within the framework of software-defined 5G and 6G networks. The main technologies, difficulties, and application possibilities for 6G-based SDN are covered in detail. The writers also covered the specifications for the cutting-edge sensors and monitoring systems that would be employed in the future 6G networks. Many crucial facets of 6G networks—such as big data and machine learning, cyber security, and industry 5.0.

The survey \cite{ porambage2021roadmap} provides a detailed discussion of security and privacy aspects of the 6G networks by visualizing various use cases; however, the authors did not provide enough discussion for the ML, AI, key technologies, and industry 5.0.

The article \cite{tang2019future} provides an overview of various machine learning methods related to the networking, transmission, and security elements of automobile networks. The authors also illustrate how artificial intelligence (AI) will be enabled for a future 6G vehicle network. Some of the topics that are illustrated include intelligent radio (IR), network intelligence, proactive exploration and self-learning, etc. High bandwidth, ultralow latency, improved reliability, and security will all be possible with AI techniques. 
 
This article \cite{yaacoub2020key} discusses the future trends in the evolution of ubiquitous connectivity in rural areas and links the requirements to the 6G networks. The authors also discussed the application and requirement aspects of the 6G network; however, the article lacks a discussion about ML and security aspects.
 
Table \ref{tab: PrevSurveys} gives a comparison of survey results from 6G. Virtual reality (VR), health, machine learning (ML), architectures, trends, and disruptive technologies are just a few of the significant and forward-thinking performance metrics that are used in this comparison. Other metrics that we have proposed include integration, security and privacy, Industry 5.0, and big data. Table \ref{tab: PrevSurveys} illustrates how most surveys fall short in areas such as industry 5.0, security, integration, and big data. Nearly all of them go into great length about ML, architectures, and difficulties. We go into great depth about each of these comparison indicators in our survey.

\begin{table*}[t]
\scriptsize
\caption{\label{tab: PrevSurveys} Summary of existing 6G surveys broken down into requirements challenges, use cases, enabling technologies, security issues, and applications of AI as well as integration with other technologies. Here, letters {\bf L}, {\bf M}, {\bf N}, and {\bf H} show less/low discussion, medium discussion, no discussion, and high discussion of the specified topic, respectively.}
  \label{tab: PrevSurveys}
  \centering
  \scalebox{0.85} {
  \begin{tabular}{c|c|c|c|c|c|c|c}
   \hline
    {\bf Survey} & {\bf Year} & \makecell{{\bf Requirements} \\ {\bf Challenges}} & \makecell{{\bf Use Cases} \\ (VR, Health, Industry5.0)} & \makecell{{\bf Enablers} \\ (Disruptive, Innovative)} & \makecell{{\bf AI} \\ (ML, Architectures, Big data, Trends)} &\makecell{ \bf  Technological \\ Integration} & \makecell{\bf {Security} \\ {\bf Privacy}}  \\ \hline
  
    \cite{gawas2015overview} & 2015 & M & (N, H, N) & (L, L)& (L, M, M, H)& L & \ L \\
    \hline
     \cite{mondal2015survey} & 2015 & M & (L, M, N) & (L, M)& (N, L, M, M)& L & \ L \\
    \hline
     \cite{vinogradov2016key} & 2016 & H & (N,M,N) & (L, M)& (H, H, L, M)& M & \ L \\
    \hline
     \cite{esmail2016outdoor} & 2016 & M & (N, M, N) & (L, L)& (M, L, N, L)& L & \ L \\
    \hline
     \cite{al2017proposed} & 2017 & M & (L, M, L) & (M, L)& (L, H, L, L)& N & \ L \\
    \hline
     \cite{ankarali2017flexible} & 2017 & M & (L, M, L) & (L, L)& (H, M, L, M)& L & \ L \\
    \hline
    \cite{yastrebova2018future} & 2018 & L & (L, M, L) & (M, L)& (L, H, L, L)& N & \ L \\
    \hline
     \cite{reka2018future} & 2018 & M & (L, M, L) & (L, L)& (H, M, L, M)& L & \ L \\
    \hline
     \cite{long2019software} & 2019& L & (L, M, L) & (M, L)& (L, H, L, L)& N & \ L \\
    \hline
     \cite{tang2019future} & 2019& M & (L, M, L) & (L, L)& (H, M, L, M)& L & \ L \\
     \hline
     \cite{yaacoub2020key} & 2020 & L & (N,M,N) & (L, M)& (H, H, L, M)& M & \ L \\
    \hline
     \cite{wang20206g} & 2020 & M & (N, M, N) & (L, M)& (M, L, N, L)& L & \ L \\
     \hline
    \cite{shahraki2021comprehensive} & 2021& H & (L, H, N) & (L, M)& (H, M, M, H)& L & \ L \\
    
    \hline
      \cite{ serghiou2022terahertz} & 2022 & L & (H. M, N) & (L, M)& (H, L, M, M)& H & \ L \\
     \hline
     \cite{ al2023edge} & 2023 & L & (H. M, N) & (L, M)& (H, M, M, M)& L & \ L \\
    \hline
     {\bf Our survey} & 2024 & H & (H, H, H) & (H, H)& (H, H, H, H)& H & \ H \\
     \bottomrule
    
  \end{tabular}
  }
 
\end{table*}

\section{{5G Applications, Standards, and Challenges for Efficient 6G}}
\label{sec:3}

 Before jumping to the 6G requirements, it is important to analyze the available generations. In the following section, we provide a brief revision of all the generations with their strengths and limitations. We also discuss the specifications and requirements for 6G networks. 

\subsection{\bf Background of Communication Generations}

Since the beginning of the first analog communication system in the early eighties, a new mobile communication generation has been launched every ten years. In the following sections, we go through these generations and present their strengths and limitations. 

{\bf First Generation.} 
First Generation (1G)  provided the service of voice calling with large network coverage. A partial fax service was also introduced in 1G networks. 1G networks faced issues like incompatibility of different market standards. For instance, at that time the USA was using an advanced mobile phone system; whereas several European countries adopted the Nordic mobile radio system (NMR). Both standards were incompatible with each other. Digital speech codecs were not introduced in the 1G era \cite {sytnikov20081g}.

{\bf Second Generation.} 
The global system for mobile communications (GSM) is considered the key standard of second-generation (2G). It began in Europe with a French name Groupe Speciale Mobile, and within years, GSM had around 90 percent of the market share. In GSM, newer voice codecs improved the voice quality. One of the key features of the GSM was the short messaging service (SMS). In addition, GSM also launched digital data services \cite{shukla2013comparative}. Initially, a data rate of 9.6 kb/s was achieved; later, the data rate was further improved to 200 kb/s with the launch of general packet radio service (GPRS) and enhanced data rates for GSM evolution (EDGE). 2G had various challenges that included low data rate, high cost, delay, etc.

{\bf Third Generation.} 
Several European regulators agreed to establish a new approach to granting licenses by the middle and end of the 1990s. It was expensive to buy a spectrum from auctions in open markets. Even after the license was granted, no standard came out. However, after several years of license granting, the universal mobile telecommunication system (UMTS)-based cell phones came onto the market. UMTS used code division multiple access (CDMA). Initially, UMTS provided data rates up to 364 kb/s which was far less than the expected rate; however, in the later modifications, higher data rates were achieved. UMTS could not provide the expected performance due to bandwidth issues. Therefore, operators decided to move towards 4G long-term evolution (LTE) \cite{ramachandran2003evolution}. 

{\bf Fourth Generation.} While working with UMTS, operators learned lessons and they were not fully confident to invest in the fourth generation (4G). Also, various technologies were introduced regarding spectrum efficiency which provided the flexibility for operators to select the appropriate option according to their requirements. Most of the LTE standards kept upgrading 3G UMTS to develop 4G communications. The LTE aimed at simplifying the existing UMTS architecture from circuit and packet switching to IP-based architecture. LTE provides up to 300 Mb/s and uploads data rates of up to 75 Mb/s. LTE offers low latency (up to 5 ms) with mobility support. It uses orthogonal frequency division multiple access (FDMA). Although, 4G provides solutions to several problems faced by 3G like low data rates and higher latency values; 4G faces other issues such as performance management for the rapid growth of the number of devices that require ubiquitous connectivity and higher data rates \cite{jaloun2010wireless}.

{\bf Fifth Generation.} 
5G is the new generation of cellular networks. According to the 3GPP project, if a device is using 5G new radio (5G NR) software, it is called a 5G device. A 5G network is supposed to provide data rates of up to 20 Gb/s. The 5G technology promises to offer a latency between 1 to 5 ms. The speed of 5G in the sub-6GHz band will be greater than the 4G with a similar number of antennas and spectrum. Overall, 5G uses a high-frequency band (above 30 GHz). In recent years, mmWave appeared as one of the potential candidates for 5G by providing data rates in gigabits/sec. It uses a spectrum between 30 and 300 GHz and corresponds to wavelengths between 10mm to 1mm. The characteristics of mmW include high bandwidth, short-wavelength/high frequency, and high attenuation \cite{denardis2020internet}. High-power levels are required to overcome the huge path loss, which is expected in mmW communications. The coverage is up to 20 meters at 60 GHz. IEEE 802.15.3c and IEEE 802.11ad standards provide support for mmW. Moreover, 5G will deliver the services like device-to-device (D2D) connectivity and massive machine-to-machine communications, etc. 5G bands have several challenges including poor foliage penetration, atmospheric and free space path loss, and high deployment cost. 

Table \ref{table: comparison} shows a performance comparison for different generations of communication systems.

\begin{table}[!th]
    \caption{An overview of characteristics of 4G, 5G, and 6G wireless networks exhibiting the advantages of 6G over 4G and 5G in terms of data rate, mobility, spectral, and network energy efficiency.}
    \begin{center}
    \label{table: comparison}
    \small
        \begin{tabular}{p{2.5cm}|p{1.5cm}|p{1.7cm}|p{1.5cm}}        \hline
            \textbf{Characteristics} & \textbf{4G}& \textbf{5G}& \textbf{6G} \\
            \hline
            Device peak data rate & 1Gbps  & 10-20 Gbps &  1Tbps  \\
            \hline
            Latency & 100 ms & 5-10 ms &  10-100 micro sec  \\
            \hline
            Maximum spectral efficiency & 15 bps/Hz & 30 bps/Hz & 100 bps/Hz  \\
            \hline
            Mobility & 350 Km/hr & 500 km/h & up to 1000 km/h  \\
              \hline
            Network energy efficiency & 1x & 10-100 x of 4G & 10-100 x of 5G  \\
            
        \hline
        
    \end{tabular}
    \label{tab1}
  \end{center}
\end{table}

\subsection{\bf Existing Industrial Standards  and Benchmark Performance Parameters for the 6G Networks}

There are several performance trade-offs in reliability, delay, throughput, and power consumption for 5G networks and these trade-offs give an impression that 5G could not fulfill the market challenges beyond 2030. The characteristics and features of 6G networks have the potential to overcome these challenges. The main features of 6G networks include high data rates, energy efficiency, reliability, and intelligent decision-making per device as well as a whole network at the global scale \cite{letaief2019roadmap}. The 6G networks are supposed to provide services including enhanced mobile broadband (eMBB),  ultra-reliable low latency communications (URLLC), massive machine-type communication (mMTC), and AI and machine learning integration. URLLC, promises to provide a delay of less than 1 ms \cite{tariq2020speculative}. Furthermore, the 6G system makes use of energy harvesting techniques so that ultra-lifetime could be provided to the devices.

For the satellite-integrated network, 6G is expected to provide improved QoS for satellite communication. 6G aims to make a single wireless operating network by integrating terrestrial and satellite networks. 6G will provide massive connectivity to wireless devices in the form of IoE; however, such huge connectivity creates challenges like quick processing and transmission delays. To cater to this, 6G will make use of extensive AI techniques at each phase of the communication including signaling, forwarding, and decision-making \cite{letaief2019roadmap}. 

Ubiquitous connectivity is considered one of the key requirements for 6G networks. In a broader aspect, drones and satellites in orbit will help to create super-3D connectivity for 6G to provide ubiquitous communication \cite{letaief2019roadmap}. 5G successfully uses the small cell communication networks to improve QoS including throughput, energy, signal quality, and spectral efficiency; however, it lacks the integration of disciplines like AI, ML, and automation \cite{chowdhury2018interference, shifat2017game, mahbas2019impact}.

6G networks will incorporate the approach of small cell networks in a larger aspect. Due to massive device connectivity, the future 6G networks will be ultra-dense with heterogeneous connectivity characteristics. 

In recent years, millimeter wave (mmW) appeared as one of the potential candidates for 5G by providing data rates up to gigabits/sec and it could be also useful for the initial implementation of 6G networks. It uses a spectrum between 30 and 300 GHz and corresponds to wavelengths between 10 mm to 1mm. The characteristics of mmW include high bandwidth, short-wavelength/high frequency, and high attenuation \cite{li2020beam, ohkoshi2020magnetic}. Huge path loss is expected in mmW communications, to recover from these path losses, high power levels are required. Due to high attenuation from solid materials (bricks and buildings), mmW requires a line of sight (LoS) for efficient and reliable communication. The interference levels in mmW communication are much lower than the 2.4 GHz and 5 GHz. Multiple-antennas solutions in mmW allow the transmission to use narrow beams which help to reduce the attenuation and the interference \cite{wang2020compressed, ju2021millimeter, he2021survey}. There have been several standardization activities for mmW MAC in the 60 GHz band. Most of these standardization efforts are for personal and local area networks under IEEE 802.15.3c and IEEE 802.11ad.
IEEE 802.15.3c standard also known as piconet specifies the mmW by supporting a high data rate of over 2 Gbps in the 60 GHz band. Among a cluster of IEEE 802.15.3c-based devices, one will be selected as the piconet coordinator (PNC) which manages the synchronization among devices by broadcasting beacon messages. The device content for the time slots using carrier sense multiple access/collision avoidance (CSMA/CA) and sending data using time division multiple access (TDMA) \cite{uwaechia2020comprehensive, xin2021low}. The IEEE 802.11ad introduced several modifications at MAC and the physical layer of the existing IEEE 802.11 standard to enable mmW support. It claims to provide a 6.75 Gbps data rate. The coordinator uses a superframe structure to manage the channel access of the connected stations which is composed of beacons, contention access period (CAP), and contention-free period (CFP) \cite{agrawal2020performance, akbar2016delay, akbar2017ieee, akbar2015holistic, akbar2016implanted}.

High-rate wireless personal area networks (WPANs) are a good fit for the IEEE 802.15.3 MAC protocol since it provides high data rates, low latency, and energy efficiency. In particular, its characteristics and capabilities support high-frequency transmission, provide quality of service, and enable sophisticated applications—all of which align with the objectives of 5G and 6G networks. IEEE 802.15.3 can greatly improve the overall performance and capabilities of future wireless networks when it is integrated into the larger 5G/6G ecosystem. It is important to understand the performance computation process for various technologies under the 5G and 6G network could be different and dependent on a specific technology. As an example, a discussion is provided regarding the performance evaluation of a 5th generation's medium access control (MAC) mechanism of the IEEE 802.15.3C standard in terms of end-to-end delay (ED) and maximum throughput (MT). The purpose is to understand how much time it takes for a channel time allocation (CTA) request in the worst-case scenario. The ED can be calculated as given in Equation~\ref{eq:1} \cite {akbar2016tmp, akbar2017ieee, akbar2018modelling}:

\begin{equation}
\label{eq:1}
ED = T_{frame} + T_{ACK} + T_{CH}
\end{equation}

Where \(T_{\mathrm{frame}}\) represents frame transmission time for a CTA request frame. Further, \(T_{\mathrm{frame}}\) can be computed as given in Equation~\ref{eq:2}:

\begin{equation}
\label{eq:2}
T_{frame} =  T_{Preamble(PHY)} + T_{Header(MAC + PHY)} + T_{Payload} 
\end{equation}
Where \(T_{\mathrm{Preamble}}\) is the duration of PLCP preamble, \(T_{\mathrm{Header}}\) is the duration of PLCP header and \(T_{\mathrm{Payload}}\) is the duration of the payload. These durations are given in the IEEE 802.15.3C standard.

\(T_{\mathrm{ACK}}\) represents the time duration of the ACK, in this case, ACK duration is computed as given in Equation~\ref{eq:3}:

\begin{equation}
\label{eq:3}
T_{ACK} =  T_{ImmACK} + 2SIFS
\end{equation}
Where \(T_{\mathrm{ImmACK}}\) is the time duration of the immediate ACK and can be computed as given in Equation~\ref{eq:4}:
\begin{equation}
\label{eq:4}
T_{ImmACK} = T_{Preamble} + T_{Header}
\end{equation}
The ACK of the Imm\-ACK has only a MAC header and not a payload as each packet is expected to be acknowledged immediately.
\(T_{\mathrm{CH}}\) represents the time to access the channel, which is computed as given in Equation~\ref{eq:5}: 
\begin{equation}
\label{eq:5}
T_{CH} = (RC \times BIFS) + (BC \times pBackoffSlot)
\end{equation}
Where RC is the  retry counter in the backoff process and its value will be 3 in the worst case as default, BIFS is the backoff IFS and it is calculated by Equation~\ref{eq:6}:

\begin{equation}
\label{eq:6}
BIFS = pSifsTime + pCcaDetectTime
\end{equation}
The values of pSifsTime and pCcaDetectTime are given in the Table I mentioned in the IEEE 802.15.3C standard. BC is the backoff counter calculated as given in Equation~\ref{eq:7}:
\begin{equation}
\label{eq:7}
BC = Rand  (0, BW )
\end{equation}
BC is computed using a random function that finds a random integer value between zero and BW (backoff window). The value of BW is given in Table \ref{table:correctivebreakdown}.

    \begin{table}[ht!]
        \begin{center}
        \caption{Timing and space parameters mentioned by IEEE 802.15.3C standard \cite{tripathy2016application}.} 
        \label{table:correctivebreakdown}
        \small
            \begin{tabular}{r|l} 
             \hline
                \textbf{PHY Parameter} & \textbf{Duration HSI ($\mu$s)} \\
                \hline
                \textbf{pSIFSTime} & {2.5} \\
                \textbf{pCcaDetectTime} & {2.5}\\
                \textbf{pBackoffSlot} & {5} \\
                \textbf{T\_Preamble} & {1.31}  \\
                 \textbf{T\_Header} & {0.44}  \\
                 \textbf{Backoff Windows} & {[7, 15, 31, 63]} \\
                \textbf{Retry Count} & { 0 to 3}  \\
                 \textbf{CAP duration ($\mu$s)} & {0 to 65,535}  \\
                  \textbf{Superframe duration ($\mu$s)} & {0 to 65,535}  \\
                 \textbf{MAC header (bytes)} & {10}  \\
                 \textbf{PHY header (bytes)} & {48}  \\
                  \textbf{Acknowledgement (bytes)} & {10}  \\
                   \textbf{Beacon packet (bytes)} & {100}  \\
                 \textbf{Data frame (bytes)} & {512 to 8,388,608}  \\
                 \textbf{Channel data rate (Gbps)} & {1.5, 3, 5}  \\
                 
             \hline
            \end{tabular}
        \end{center}
    \end{table}

 The maximum throughput (MT) is defined as the ratio of transmitted information in bits to the transmission duration. Throughput is defined as the ratio of payload size ($X$) 
 to the total time required to transmit the payload, in the case when there is no priority set the maximum network throughput can be computed as given in Equation~\ref{eq:8}: 
\begin{equation}
\label{eq:8}
MT = \frac{X}{ED}
\end{equation}
In this regard, multi-tier network architecture will help to manage these ultra-dense networks in 6G. To manage the huge traffic in 6G networks, high-capacity backhaul networks are required \cite{akyildiz2014terahertz}. This can be possible by using free-space optical (FSO) systems and a fast optical fiber network. Overall, 6G wireless networks will be able to:
\begin{itemize}
    \item Provide super-high-definition (SHD) and extremely high-definition (EHD) video transmission which requires ultra-high throughput.
    \item Support extremely low latency up to 10 microseconds.
    \item Manage the Internet of Nano-Things and Internet of Bodies using smart implantable and wearable devices under very low power consumption.
    \item Enable efficient underwater and space transmissions to drastically increase the limits of human activity, for example, deep-sea exploration and space traveling.
    \item Support various advanced service practices like hyper-high-speed railway (HSR).
    \item Improve 5G applications, like the huge Internet of Things (IoT) 
    and entirely autonomous vehicles.
    \item Support enhanced mobile broadband communications, ultra-huge machine-type communications, enhanced ultra-reliable and low-latency communications, and long-distance and high-mobility communications.
  \end{itemize}    

\subsection{\bf Regulatory Requirements}

The following are some potential regulatory requirements that may be relevant for 6G networks:
{\bf Spectrum Allocation.}
 Regulatory bodies need to allocate sufficient spectrum resources for 6G networks to operate. This may involve identifying new frequency bands or optimizing the use of existing bands to accommodate the higher data rates and increased capacity demands of 6G.

{\bf Licensing and Spectrum Management.}  Governments may establish licensing frameworks and spectrum management policies to ensure fair access to spectrum resources. This may include licensing conditions, auction processes, and spectrum-sharing mechanisms to promote competition and efficient spectrum utilization.

{\bf Security and Privacy.} 
Regulatory requirements related to security and privacy are likely to be strengthened for 6G networks. This may involve mandating robust encryption standards, authentication mechanisms, and privacy safeguards to protect user data and ensure network integrity.

{\bf Quality of Service (QoS).} Regulatory bodies may define minimum QoS requirements to ensure reliable and consistent service delivery in 6G networks. This may include criteria for latency, throughput, reliability, and other performance metrics.

{\bf Interoperability and Standards.} Regulatory frameworks may encourage the adoption of common technical standards and interoperability among 6G networks and devices. This promotes seamless connectivity, and interoperability between different vendors, and avoids fragmentation in the network ecosystem.

{\bf Environmental and Health Considerations} As with previous generations of wireless networks, 6G may be subject to regulations addressing potential environmental and health concerns. This may involve assessing and mitigating the impact of radiofrequency emissions and ensuring compliance with safety guidelines.

\subsection{\bf Other Issues} 
Data storage issues, resource restrictions on edge devices, heterogeneous device management, and performance QoS issues are potential obstacles for 6G networks. Here are a few instances:

{\bf Data management.} 
Managing the enormous volumes of data that will be produced by the growing number of connected devices will be one of the issues faced by 6G networks. The Internet of Things (IoT) will bring forth an influx of data that needs to be processed, stored, and analyzed. This will need for brand-new data management and storage strategies that can scale to meet the needs of 6G networks. For instance, edge computing and distributed storage systems can ease the strain of centralized data centers by moving computation closer to the source of data generation.

{\bf Resource restrictions on edge devices.} 
Managing resource limitations on edge devices will be a difficulty for 6G networks. Smartphones, wearables, and IoT devices will have to function with constrained processing speed, memory, and battery life. To do this, new methods of resource optimization, such as edge caching, network slicing, and adaptive resource allocation, will be needed. To save battery life and limit the need for network connectivity, edge caching, for instance, can be used to save frequently accessed data locally on the device.

{\bf Management of heterogeneous devices.}
6G networks will also need to handle a variety of heterogeneous devices with various capabilities and needs. To achieve interoperability and support device diversity, new methods of device management will be necessary. For managing devices with various operating systems and hardware architectures, for instance, device virtualization and containerization can offer a uniform platform.

{\bf Performance and QoS.}
To support applications with strict performance requirements, such as AR, VR, and autonomous vehicles, 6G networks will need to offer high levels of QoS. This calls for innovative methods for network optimization and scheduling that can give traffic the attention it deserves based on its QoS specifications. Network slicing, for instance, can be used to designate specific network slices for applications with various QoS requirements.

\section{6G Enabling Technologies and AI Role }
\label{sec:4}

This section describes the technological enablers that are supposed to provide the 6G revolution. These key enablers will justify the migration from 5G to 6G networks. We discuss various technologies including Terahertz (THz), optical wireless technology, full-duplex communication, channel estimation techniques, network localization, MIMO, 3D network architecture, network virtualization, FSO, energy harvesting techniques, blockchain, WIET, and supervised and unsupervised learning \cite {jornet2011channel, komine2004fundamental, goyal2015full, ali2017millimeter, schloemann2015toward, wang2017machine, akyildiz2014terahertz, tekbiyik2019terahertz, siaudthz, gu2018network, douik2016hybrid, alghamdi2020intelligent}.  Based on technological characteristics, we categorize enabling technologies into three main categories: { disruptive, innovative, and intelligence integration}. We discuss these technologies in the following sections.

Table~\ref{tab: AIMech} provides the details of various AI mechanisms with their applications. Recently, a variety of AI-based mobile applications emerged that require stringent QoS communication services. Many works have tried to address these requirements  \cite{wollschlaeger2017future, tariq2020speculative, chowdhury2018interference, shifat2017game, mahbas2019impact, aste2017blockchain, miller2018blockchain, dai2019blockchain, wang2014wireless, nawaz2019quantum, elsayed2019ai}. 

\begin{table*}[!h]
\centering
\caption{\label{table: SecuritySummary} Security and privacy issues in 6G-enabling technologies. 
}

\small
  \begin{tabular}{p{2.5cm}|p{1.5cm}|p{2.5cm}|p{9.5cm}} \hline
    \textbf{Enabling Technology} & \textbf{Paper} & \textbf{Security Issues} & \textbf{Proposed Solution} \\ 
    \hline
     Terahertz & \cite{akyildiz2014terahertz} & Authentication & Electromagnetic signatures are proposed  \\
    \hline
   Terahertz & \cite{ma2018security} & Malicious activity & The authors highlighted that signal is transmitted via a narrow beam, it can still be intercepted by an eavesdropper  \\
    \hline
    Artificial Intelligence & \cite{loven2019edgeai} & Access control & Authors proposed fine-grained control processes \\
    \hline
    Artificial Intelligence & \cite{mahdi20215g} & Malicious activity & Proposed anomaly detection system  \\
    \hline
    Artificial Intelligence & \cite{sattiraju2019ai} & Authentication & Unsupervised learning process is used as solution  \\
    \hline
    Artificial Intelligence & \cite{hong2019machine} & Communication & Antenna-based communication is mapped with AI algorithm as a secure solution  \\
    \hline
   Artificial Intelligence & \cite{nawaz2019quantum} & Encryption& ML and quantum-based encryption schemes  \\
    \hline
    Blockchain & \cite{kiyomoto2017blockchain} & Authentication & A novel architecture is proposed for authentication   \\
    \hline
    Blockchain & \cite{kotobi2018secure} & Access control & Proposed a method that improves access control  \\
    \hline
    Blockchain & \cite{ferraro2018distributed} & communication & Proposed to use hashing mechanism to validate transactions  \\
    \hline
    
  \end{tabular}
\end{table*}

This section discusses how 6G will manage mobile AI applications.
\subsection{\bf Trends and Challenges}
AI has performed well in many domains including computer vision and natural language processing. AI processes are computationally rigorous and implemented at data centers with custom-designed servers. With the advancement of mobile communication and Internet of Things devices,  numerous applications will be operating in the future and are expected to be implemented at the edge of wireless networks. The design of the 6G wireless network needs to consider the hardware, QoS, and software requirements for IoT-based applications. Wireless networks in the 6G-AI context face various challenges including privacy, data storage, resource limitations on edge devices, managing heterogeneous devices, and performance QoS \cite{mcmahan2017communication}. AI techniques such as federated learning, deep learning, and neural learning have the potential to meet the challenges of 6G wireless networks \cite{luong2019applications}.

\subsection{\bf Security and Privacy}
\label{subsec:security}
Security and privacy are paramount concerns in the development of 6G networks, particularly given the complexity and novelty of the technologies involved. In this subsection, we aim to delve deeper into these critical issues, which are crucial for ensuring the robustness and reliability of 6G networks across various applications, including healthcare sensing devices and industrial automation.

The security and privacy challenges encompass several key aspects such as access control, detection of malicious activities, authentication, and encryption, all of which are fundamental to safeguarding sensitive data and ensuring the integrity of communications. These issues are discussed comprehensively about the enabling technologies discussed earlier in this paper.

Table \ref{table: SecuritySummary} provides an overview mapping these security and privacy concerns with specific enabling technologies such as Terahertz, AI, and blockchain. For instance, in Terahertz communications, authentication mechanisms based on electromagnetic signatures have been proposed to mitigate security risks \cite{ma2018security}. Similarly, in the context of AI applications within 6G networks, access control mechanisms, malicious activity detection algorithms, and robust authentication protocols have been developed to enhance security \cite{loven2019edgeai, mahdi20215g, sattiraju2019ai, hong2019machine, nawaz2019quantum}.

Furthermore, blockchain technology offers promising solutions for ensuring secure transactions and communications in 6G networks through innovations in authentication and decentralized access control \cite{kiyomoto2017blockchain, kotobi2018secure, ferraro2018distributed}.

Despite these advancements, it is acknowledged that the discussion provided here requires further depth and critical analysis. Future research and development efforts should focus on addressing the inherent challenges and complexities of implementing robust security and privacy measures in 6G networks. This includes not only technical considerations but also regulatory and ethical implications to ensure comprehensive protection against emerging threats. While significant strides have been made in addressing security and privacy concerns in the context of 6G, ongoing efforts are essential to fully harness the potential of these technologies while mitigating associated risks effectively.

\subsection{\bf Communication for Distributed Machine Learning}
Along with AI techniques, high-level distributed machine learning techniques are required to make a successful execution of wireless networks in a 6G environment. The machine learning techniques elevate the complexities of the cloud wireless architecture, decision-making at the network edge, and end devices.

{\bf Communication-Efficient Distributed Training.} 
The expanding processing and storage capacity of advanced communication makes devices capable of on-device training, learning, and processing of received data. Yet, transmitting over the unpredictable wireless channel turns into a substantial performance bottleneck for mobile distributed networks. To handle the security and privacy issues, federated learning \cite{kato2019optimizing}  keeps the training data at each device. Federated learning is an emerging distributed ML/AI approach with privacy preservation, is specifically useful for various wireless applications, and is used as one of the key solutions to obtain ubiquitous AI in 6G.

{\bf Communication-Efficient Distributed Inference.} In 6G networks, it is expected that AI services will be implemented at data centers and end IoT devices, e.g., unmanned vehicles and drones. A lot of research efforts have been made to design inference processes with very low latency and cost. To handle issues like massive computation, processing, power, and privacy in 6G networks, mobile edge computing is considered a potential candidate \cite{luong2019applications}. For instance, in the case of a deep neural network, at end devices, the features can be extracted for decision-making, and later this information can be communicated to the edge/cloud computed devices. However, the heterogeneous nature of devices in  6G networks makes computing and inter-processing mechanisms more challenging for neural networks \cite{mcmahan2017communication}.

\begin{table*}[htbp]
\centering
    \caption{Communication layers and AI algorithms (listed and detailed in Table~\ref{table: new}). 
    }
    \begin{center}
    \label{table: layers}
        \small
        \begin{tabular}{p{2.5cm}|p{9.5cm}|p{4.5cm}}
        \hline
            \textbf{Network Functions} &  \textbf{Descriptions}  & \textbf{AI Algorithms}\\
            \hline

                   Physical Layer
                 
                  &  
                The physical layer transmits the data on the wireless medium using different modulating schemes
                including channel coding, advanced modulation schemes like OFDM and MIMO, etc.
                 & 
                 K-Means, DNN, CNNs, CCNNs, Auto-encoder\\
             \hline
                  Data Link Layer  
              &  
                 The data link layer is mainly responsible for the frame Performance issues i.e., scheduling, 
                error identification and correction mechanisms, flow control mechanisms, synchronization,  queuing, etc.,
                    
                     & 
                 DNNs, Q-Learning, Reinforcement Learning, Supervised Learning, Transfer Learning\\
             \hline
                  Network Layer

              &  
                The network layer is responsible for connection management, mobility management,
                load and routing protocols management
                     & 
                DNNs, Reinforcement Learning, Supervised Learning, Unsupervised Learning, K-Means, Q-Learning\\
                 
                
                
                
                
        \hline
        
    \end{tabular}
    \label{tab1}
  \end{center}
\end{table*}

\subsection{\bf Disruptive Communication Technologies} 
Communication technologies like multiple-input-multiple-output (MIMO) and mmWave are considered key enablers of 5G networks and provide the required QoS. To meet the requirements of 6G networks, 6G will exploit the unused terahertz band and visible light communications (VLC).

{\bf Terahertz Communications.}  
The radio frequency (RF) band is nearly exhausted and not enough to fulfill the 6G requirements. Spectral efficiency can be enhanced by expanding the bandwidth, which can be achieved by applying THz bands and massive MIMO technologies. The THz band will play an important role in 6G communication \cite{ tekbiyik2019terahertz}. The THz band is intended to be the next frontier of high-data-rate communications. THz communications use frequency bands between 100 GHz and 10 THz with wavelengths in the 0.03 mm – 3 mm range. ITU-R recommends 275 GHz - 3 THz bands for the communication of cellular networks \cite{siaudthz}. In comparison with mmWave, THz faces challenges due to ultra-high frequencies. The main challenges that are hurdles for terahertz’s commercial adoption are propagation loss, molecular absorption at a higher frequency, the high penetration loss for solid objects, and, the antenna design complexity. In mmWave, the propagation loss can be controlled using the directional antenna arrays mechanism. However, some of the frequencies in the terahertz spectrum are more sensitive to atmospheric molecular absorption. This issue can be avoided by only using selected frequency bands in the Terahertz spectrum  \cite{jornet2011channel}.

{\bf Optical Wireless Technology.} 
Optical wireless technology (OWC) is considered a key enabler for 6G communications; however, it is already being used partially in 4G/5G networks. Main OWC technologies include light fidelity, optical camera communication, VLC,  and FSO communication \cite{chowdhury2018comparative, chowdhury2019integrated, hossan2018new, hossan2019human}. Networks with OWC technologies have the potential to provide ultra-throughput and very low latency. VLC is proposed to support RF communications using light-emitting diode (LED) luminaries. These devices can switch among various light intensities to modulate the signal. The research activities on VLC are more progressive than the terahertz communications. Although, the IEEE 802.15.7 standard is proposed for VLC, yet not been adopted by 3GPP. VLC has a short range, requires an illumination source for communication, and is also sensitive to the noise created by any light source including the sun \cite{komine2004fundamental}. Due to these reasons, it is mostly recommended for indoor environments.


{\bf Full-Duplex Communication.} Recently, cellular networks have adopted full-duplex solutions which will make the base station’s transceiver capable of transmitting a signal while receiving within a limited range, such design is achievable using innovative design of self-interference suppression circuits \cite{goyal2015full}. The main innovation is being achieved by improving the design of the antenna and circuit which reduces the cross-talk between receiver and transmitter for a wireless device \cite{hausmair2017digital}.

In the future, technological advancement will be capable of simultaneous transmission and reception. Such advancements enhance the multiplexing capabilities of the existing cellular systems in terms of throughput. A 6G network could be very useful for such a full-duplex network in terms of performance requirements.

{\bf Novel Channel Estimation Techniques.} 
For cellular networks, channel estimation for beam tracking is considered an important element for ultra-high frequencies. Therefore, advanced channel estimation techniques will be highly beneficial in cellular networks. If we talk about mmWave, design challenges exist for channel estimation techniques due to complex antenna architectures. A recent research effort has been made to provide channel estimation for mmWave, known as out-of-band estimation, which provides an improvement in the reactiveness of beam management schemes  \cite{ali2017millimeter}.

{\bf Sensing and Network-Based Localization.} Techniques based on RF signals have been broadly used to empower concurrent operation of localization and mapping techniques  \cite{zanella2014internet}. However, it is still not adopted for the case of cellular networks. In this regard, 6G will provide an integrated interface to support localization and mapping mechanisms which will offer multiple advantages including enhancement in control operations, less interference, and novel services for applications like eHealth. 

{\bf Advanced Multiple-Input Multiple-Output.} Multiple antenna technologies have achieved considerable response from both academia and industry as they are capable of providing high geographical coverage and spatial multiplexing which helps to boost spectral efficiency \cite{gao2019antenna, attarifar2019modified} and could be very useful to achieve the performance goals of 6G networks. Therefore MIMO is a potential area that can integrate itself with 6G networks which will increase the capability of 6G network devices in terms of transmission. 

\subsection{\bf Innovative Network Architectures}
{
The 6G network offers various innovations to the existing communication architecture, such as virtualization. Significant developments in wireless communication technology are being driven by the future implementation of 6G networks and the transition to 5G. Among these developments, the terahertz (THz) and millimeter-wave (mmWave) frequencies are proving to be crucial facilitators due to their capacity to accommodate extremely high data rates and extensive communication. Two key elements of this architecture are the provision of numerous network services and coordinated multiuser beamforming training. This document describes the system architecture for these capabilities, with an emphasis on the integration and optimization of THz/mmWave technologies within the framework of 5G and 6G networks. The 6G network architecture includes multiple components, such as Base Stations (BS) and Access Points (AP), User Equipment (UE), Beamforming Modules, Network Core, Beam Training Protocols, and Multiple Network Services Framework \cite{hausmair2017digital}. }

In the following, we discuss some of the promising architectural domains and examine all the main components.

{\bf Cell-Less Architecture and Tight Integration of Multiple Frequencies}
The 6G network will create a strong relationship between cells and the user equipment (UEs) and will create a cell-less architecture. This can be accomplished using dual-band techniques with multiple radios. Such architectures provide high mobility support with fewer overheads. It will increase the user capacity by seamlessly transitioning to multiple heterogeneous links (e.g., to identify the best channel) without any prior configurations \cite{ serghiou2022terahertz}. 

{\bf Ultra-Massive MIMO (Um-MIMO).}
A major increase in antenna count is incorporated into Um-MIMO, an advancement of MIMO technology that improves spectrum efficiency, capacity, and reliability. Because it allows for the integration of a large number of devices and the transmission of high data rates, this technology is essential for 6G networks. Recent developments concentrate on enhancing beamforming methods and lowering interference, which makes Um-MIMO a crucial enabler for high throughput and low latency applications like virtual reality (VR) and augmented reality (AR).

{\bf Radio Resource Management (RRM).} 
To satisfy the varied needs of different applications and services, RRM in 6G networks involves dynamic allocation and optimization of radio resources. RRM incorporates artificial intelligence and machine learning techniques to improve decision-making processes that will lead to better quality of service (QoS) and more effective spectrum utilization. These developments are crucial for controlling the higher demand and complexity in 6G networks, especially in situations where there is a high level of mobility and a dense population of devices. 

{\bf Non-Orthogonal Multiple Access (NOMA).} 
NOMA dramatically increases spectrum efficiency by leveraging power domain separation to enable several users to share the same frequency resources. By allowing more users to share the same bandwidth, NOMA is anticipated to provide huge connections and high data rate transmission in 6G networks. This is very helpful for Internet of Things applications and other situations where a lot of devices need to talk to each other at once. To improve NOMA's performance in 6G, research is still focused on power allocation and interference management optimization \cite{vinogradov2016key}.

{\bf Coordinated Multi-Point (CoMP).}
CoMP allows numerous base stations to coordinate their receptions and transmissions, improving network performance. At the cell borders, this synchronization enhances signal quality and lessens inter-cell interference. CoMP plays a key role in enabling ultra-reliable and low-latency communications (URLLC) in 6G networks, which are necessary for applications like remote surgery and autonomous driving. Research endeavors to enhance coordination mechanisms and create more advanced algorithms to optimize the advantages of CoMP in intricate network settings \cite{ al2023edge}.

{\bf Rate-Splitting Multiple Access (RSMA).}
By splitting the data stream into common and private sections, RSMA, a cutting-edge multiple-access technique, enables more effective and flexible resource allocation. This technique can improve communications reliability, reduce latency, and increase spectrum efficiency, making it viable for 6G networks. When it comes to managing the varied and dynamic traffic patterns in 6G, RSMA is anticipated to be crucial to meeting the network's performance objectives \cite{vinogradov2016key}.

{\bf 3D Network Architecture.} 6G networks will support three-dimensional (3D) visualization by using heterogeneous architecture. Communication technologies like drones and satellites will use these architectures to improve related QoS. Furthermore, these heterogeneous communication architectures could be easily deployed in rural and disaster areas \cite{xu20113d}. 

{\bf Virtualization of Networking Equipment.} In 6G networks, network virtualization will be one of the key concepts. The virtualization will be implemented at different layers i.e., the network layer, MAC layer, and physical layer. It will reduce the management burden from hardware devices cost-effectively. It will create a centralized virtual controller which will be responsible for the policy and data forwarding \cite{alam2020survey}.

{\bf Advanced Access-Backhaul Network (FSO) Integration.} In 6G networks, technologies are supposed to provide ultra-high data rates which also require sufficient expansion of backhaul capacity. Furthermore, future VLC and terahertz implementations will use a massive number of devices, which need backhaul connectivity with the core network and neighboring nodes  \cite{schloemann2015toward}. 

\begin{figure}[htb!]
        \centering
            \includegraphics[width=0.5\textwidth]{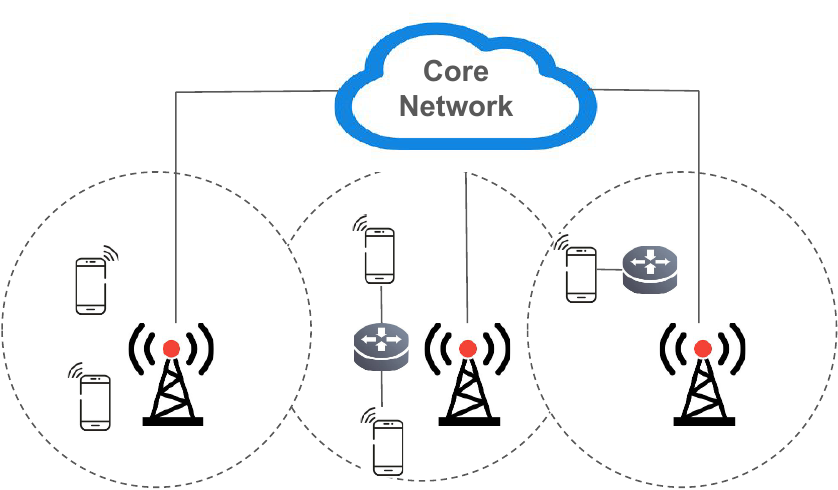}
            \caption{An overview of Integrated backhaul link \cite{deva2021dynamic}. Here, the ``Core Network'' interconnects all three cell towers, demonstrating how mobile devices communicate with the core network infrastructure via linked backhaul connections. }
            \label{Figure:BH}
\end{figure}

In some cases, such as, at remote geographical locations  (e.g., sea, space, and underwater), it is not feasible to deploy optical fiber as a backhaul network. For such scenarios,  the FSO backhaul network is a potential communication system for 6G networks \cite{gu2018network, douik2016hybrid, bag2018performance}. FSO performs like an optical fiber network and provides similar communication services. FSO can provide a large communication range, even greater than 10,000 km. Figure \ref{Figure:BH} shows an example of an integrated backhaul link.

{\bf Energy Harvesting Strategies for Low-Power Consumption Network.}   
Energy/power harvesting is defined as the process by which energy is extracted from external resources including solar, thermal, wind, and kinetic energy, etc., and then preserved for wireless autonomous devices, like wearable biomedical monitoring sensors. The energy harvesting process is capable of producing a small amount of energy which is an ongoing challenge for the industry. One of the oldest applications of energy derived from ambient electromagnetic radiation (EMR) is known as crystal radio \cite{pei2021review}. 

5G networks do consider energy harvesting techniques; however, not much work is done on that due to a lack of interest from manufacturers. As 6G networks deal with a massive number of interconnected smart devices (e.g., IoT devices ) that require energy resources to keep the device alive. There is a need to identify suitable energy-efficient mechanisms for such devices \cite{wang2017machine}.  

{\bf Blockchain.} Blockchain manages huge datasets in an efficient way \cite{henry2018blockchain, aste2017blockchain, miller2018blockchain}. Blockchain works using the approach of distributed ledger technology. A distributed ledger’s database operates in a distributed manner by using various nodes. Every node replicates a copy of the ledger. The nodes can work independently without involving the centralized node. In the blockchain, the data is managed in the form of blocks that are connected securely. The blockchain can manage huge data involved in IoTs-based systems \cite {dai2019blockchain}. Hence, 6G capabilities will support blockchain-based applications in terms of their performance goal.

{\bf Integration of Wireless Information and Energy Transfer (WIET).} Recently, producing energy from the open environment is one of the potential areas that can help to increase the lifetime of battery-operated wireless sensor devices. WIET is recognized as one of the innovative technologies for 6G networks. WIET also utilizes the wireless spectrum for communication. WEIT can be very useful for battery-less devices which are supposed to operate in 6G networks. The base stations in 6G will be used for transferring power as Wireless Information and Energy Transfer (WIET) uses the same fields and waves used in communication systems. WIET is an innovative technology that will allow the development of battery-less smart devices, charging wireless networks, and saving the battery life of other devices \cite{elmeadawy2020enabling, wang2014wireless}.

{\bf Open Radio Access Network (O-RAN).} It is an architecture that has emerged as a promising candidate for next-generation 6G networks~\cite{chen2024evolution}. O-RAN seeks to promote openness, intelligence, and flexibility in the radio access network (RAN) by leveraging standardized interfaces and modular components~\cite{tip_openran,oran_alliance}. This approach facilitates the integration of multi-vendor equipment and advanced technologies, driving innovation and reducing deployment costs. O-RAN is built on several foundational principles that involve the adoption of standardized open interfaces that allow the interoperability of equipment from different vendors. This openness helps break the traditional dependency on single-vendor solutions, fostering a more competitive and innovative ecosystem. By encouraging a multi-vendor ecosystem, O-RAN reduces costs and enhances innovation~\cite{tip_openran}. The modular design and open interfaces of O-RAN enable more flexible and efficient network deployments and upgrades. Furthermore, the integration of AI and ML supports advanced network management capabilities, improving overall performance and user experience. Achieving seamless interoperability across different vendors' equipment requires rigorous standardization and testing. Additionally, the openness and modularity of O-RAN introduce new security challenges, discussed in Section~\ref{subsec:security}, that must be addressed to ensure robust protection against cyber threats~\cite{liyanage2022openransecuritychallenges,oran_sec}.

\subsection{\bf Integrating Intelligence in the Network}
The 6G network complexity in terms of architecture and data requires AI and machine learning techniques to play their role in the successful execution of 6G services. Although 6G does not specify specific techniques, more likely AI based on a data-driven model could be used for the successful deployment of 6G networks \cite{wang2017machine}. Currently, 5G networks only integrate partial AI and ML techniques while it is expected that in 6G networks AI and ML will be fully integrated with 6G networks like the concept of Industry 5.0.  

{\bf Unsupervised and Reinforcement Learning.} Unsupervised and reinforcement learning techniques could be used as a potential learning technique in 6G networks. In 6G networks, a massive amount of data will be generated and it will not be an easy task to assign labels as required in supervised learning; whereas, unsupervised learning does not require such labeling and is capable of representing a learning model for such data. Furthermore, unsupervised learning in combination with reinforcement learning could produce efficient autonomous systems \cite{letaief2019roadmap}.

\begin{table*}[htb!]
    \caption{\label{table: Architecture} Breakdown of 4G, 5G, and 6G wireless networks' features across usage scenarios, applications, network characteristics, and relevant technologies. For acronyms, please refer to Table~\ref{tab:acronyms}. }
    
    \begin{center}
     \small   
        \begin{tabular}{p{2.75cm}|p{3.15cm}|p{4.5cm}|p{6cm}}
        \hline
            \textbf{Features} &  \textbf{4G}  & \textbf{5G} & \textbf{6G}\\
            \hline

               Usage Scenarios
                 
                  &  
                  
                 MBB
                    
                     & 
                     eMBB, URLLC, mMTC
                  
                  
                  
                    & 
                    umMTC, ERLLC, ELPCs, LDHMC, FeMBB\\
                  
                 
             \hline
               Applications                 
               &  High-Definition Videos, Voice, Mobile TV, Mobile Internet, Mobile Pay
                     &
                     VR/AR/360$^{\circ}$ Videos, UHD Videos, V2X, IoTs, Smart City, Telemedicine, Wearable Devices
                  
                    & 
                umMTC, Holographic Verticals and Society, Tactile/Haptic Internet, Full-Sensory Digital Sensing and Reality, Fully Automated Driving, Industrial Internet, Space Travel, Deep-Sea Sightseeing, Internet of Bio-Nano-Things\\
                  
 
                    
            \hline
            
               Network Characteristics
                 
                  &  
                  Flat and All-IP based
                     & 
                Intelligence integration, Clouds architecture, Software virtualization, Slicing
                    & 
               Cloudization, Softwarization, Virtualization, Slicing
                    \\        
             \hline
               Technologies
                 
                  &  
                OFDM, MIMO, Turbo Code, Carrier Aggregation, Hetnet, ICIC, D2D Communications, Unlicensed Spectrum                    
                     & 
                mm-wave Communications, Massive MIMO, LDPC and Polar Codes, Flexible Frame Structure, Ultradense Networks, NOMA, Cloud Computing, Fog/Edge Computing, SDN/NFV
                    & 
    THz Communications, SM-MIMO, LIS and HBF, OAM Multiplexing, Laser and VLC, Blockchain-Based Spectrum Sharing, Quantum Communications and
                Computing, AI/Machine Learning
                    \\
                    
        \hline
        
    \end{tabular}
    \label{tab1}
  \end{center}
\end{table*}

\subsection {\bf AI-Enabled Technologies for 6G Wireless Networks}\label{sec:5}

AI technologies have the potential to be extremely important in addressing several issues related to the installation and functioning of 6G networks. The rapid advancement of wireless networks will make 6G significantly distinct from the earlier generations, as it will be exemplified by a high level of heterogeneity in several aspects including network infrastructures, network access mechanisms, dual-band devices, processing resources, etc. Furthermore,  the size of the massive data generated in wireless networks is rising substantially. In this section, we promote AI algorithms (detailed in Table~\ref{table: new}) as an essential tool to expedite intelligent decision-making and learning for 6G wireless networks \cite{kato2019optimizing}. We discuss and review different works which addressed these issues \cite{bag2018performance, gao2019antenna, attarifar2019modified, henry2018blockchain, luong2019applications, mcmahan2017communication, kato2019optimizing,chowdhury2018comparative, chowdhury2019integrated, hossan2018new, hossan2019human, gu2018network, elliott2019recent, carmigniani2011augmented}.


{\bf Big Data Analytics for 6G.} 
For 6G wireless networks, four main categories of data analytics can be potentially applied. These categories include descriptive, diagnostic, predictive, and prescriptive analytics. Descriptive
analytics uses historical data sets to obtain perceptions performance indicators including traffic profiles, channel states, user viewpoints, etc.  Descriptive analysis helps to extend the familiarity of the network operators \cite {loven2019edgeai, mahdi20215g, sattiraju2019ai}.  
Diagnostic analytics is used for the identification of network failures with the potential reason for failure or fault so that future systems can be improved.
Predictive analytics closely observes the data pattern in historical data and predicts future behavior and predictions. Predictive analytics are tied with machine learning techniques to get optimal results. Prescriptive analytics helps to improve various domains in learning including resource allocation, network virtualization, cache management, edge, fog computing, etc \cite{mcmahan2017communication}.

{\bf AI-Enabled Closed-Loop Optimization.} Existing optimization mechanisms for wireless networks are considered to be inappropriate for 6G wireless networks due to various characteristics of 6G including dynamicity, complexity, the massive density of devices, and heterogeneity. Furthermore, existing mathematical optimization techniques have shortcomings for such massive 6G networks. For example, the objective functions in the algebraic form help optimizers and make a simple solution; however, 6G wireless networks have different scenarios, and these solutions are not suitable.
One of the potential AI mechanisms is an automated and closed loop that can be used for 6G wireless networks. In this regard, current AI mechanisms, like reinforcement and deep reinforcement learning (DRL) can be used, which can establish a feedback loop between the decision process and the wireless network system. By using this approach, the decision-making process will make the best decision based on the current situation \cite{luong2019applications}.

{\bf Efficiency and Spectrum Management.} 
Managing the increasingly crowded spectrum effectively, especially with the introduction of new frequency bands such as mmWave is a huge challenge for 6G networks. In this context, Real-time usage patterns can be analyzed and spectral resource distribution can be optimized by AI-driven dynamic spectrum management. By predicting spectrum demand and dynamically allocating frequencies to various users and applications, machine learning algorithms can improve spectral efficiency.

{\bf Network Optimization and Self-Organization.} 
Another difficult task is handling the complexity of heterogeneous 6G networks, which will have a variety of frequency bands, several access protocols, and a large number of linked devices. Self-organizing networks (SONs) with AI capabilities can automate tasks including network optimization, configuration, and repair. To guarantee smooth connectivity and performance, machine learning models can optimize factors like power levels, antenna tilts, and handover thresholds by analyzing network data and predicting and mitigating possible problems.

{\bf Latency Reduction and Reliability.} 
6G faces challenges in achieving ultra-low latency and excellent dependability, which are essential for applications like industrial automation, remote surgery, and autonomous driving. AI can forecast traffic trends and instantly modify network resources to reduce latency. When AI and edge computing are used together, latency can be greatly decreased by processing data closer to the end consumers. AI can also improve mistake detection and correction systems, guaranteeing improved data transmission dependability.
 
{\bf Energy Efficiency.} 
It is a difficult issue to reduce the energy consumption of 6G networks, which will consist of a large number of base stations and connected devices. By dynamically modifying the power consumption of network nodes in response to actual demand, artificial intelligence (AI) can optimize energy usage. Machine learning algorithms can recognize trends in network utilization and forecast times of low traffic, which makes it possible to implement energy-saving techniques like turning down specific base stations or lowering their power consumption during these periods.

{\bf Self-healing in Wireless Networks.} 
Advanced AI technologies could be useful to optimize the physical layer’s performance of wireless networks. The existing wireless systems have various physical layer issues including noise, channel and hardware impairments, quadrature imbalance, interference, and fading. AI technologies can provide an optimal way to communicate among different hardware. In the future, AI is supposed to provide self-healing and self-optimizing mechanisms for sensor-based 6G networks \cite{reshmi2021improved}. 

\subsection{\bf Summary}

Table \ref{table: layers} shows the relation between various network layers and potential AI algorithms for those layers. AI algorithms (detailed in Table~\ref{table: new}) can play a critical role in improving the performance of operational activities of the network model layers. AI algorithms will somehow automate the functionality of these layers. For the physical layer, K-Means, DNNs, CNNs, and CCNSs algorithms are suggested. These algorithms can easily be mapped with physical layers modulating techniques like OFDM and MIMO. To optimize the frame performance in terms of scheduling and error detection, at the data link layer, Q-learning, supervised learning, and reinforcement learning are suggested. At the network layer, to increase the performance for routing and mobility,  reinforcement learning, supervised and unsupervised learning, K-Mean, and Q-Leaning are advised. 

\begin{table*}[t]
    \caption{Network Architectures and AI Algorithms.  }
    \label{table: AI}
    \begin{center}
     \small   
        \begin{tabular}{p{4cm}|p{7cm}|p{5cm}}
        \hline
            \textbf{Network Functions} &  \textbf{Descriptions}  & \textbf{AI Algorithms}\\
            \hline

                   Software Defined Networks (SDN)
                 
                  &  
                In SDN, the control and data-forward
                function is decoupled to achieve programmable
                network management and configuration
                    
                     & DNN, Enhanced Q-Learning, Support Vector Machines, Self-Organizing Maps,
                  Biological Danger Theory,
                  Gradient-Boosted Regression,
                  Deep Reinforcement Learning,
                  
                    \\
                 
             \hline

                  Network functions virtualization (NFV)

              &  
                 In NFV, hardware is decoupled from Software
                and reduce the dependency of
                network functions over hardware
                    
                     & Self-Organizing Maps,
                  Biological Danger Theory,
                 Gradient-Boosted Regression,
                 Deep Reinforcement Learning
                  
                  
                    \\
                 
             \hline

                  Cloud/Fog/Edge Computing

              &  
                 Cloud computing represents a large pool of servers and systems that Provides data access and uses networks of shared IT architecture.
                Fog computing extends the edge of the network and enhances the operation of computing, resources, and services between end devices and data centers over cloud architectures.
                Edge computing brings processing near to the data centers which improves the performance of data transmission.
                
                     &  
                 Enhanced Q-Learning,
                 Support Vector Machines, 
                 Self-Organizing Maps,
                 Biological Danger Theory,
                 Gradient-Boosted Regression,
                  Deep Reinforcement Learning
                  
                    \\

        \hline
        
    \end{tabular}
    \label{tab1}
  \end{center}
\end{table*}

 Table \ref{table: Architecture} provides a comparison of 4G, 5G, and 6G technologies in terms of usage scenarios, applications, network characteristics, and technologies. In 4G networks, a simple implementation of MBB was available which evolved into eMBB and provides better mobile broadband. 5G networks also support eMBB, mMTC, and URLLC and it is expected that these technologies with further enhanced like umMTC and ERLLC will support enhanced low latency and massive machine-to-machine communication. The use of ERLLC, umMTC, and FeMBB 6G will support innovative applications like Holographic verticals, Internet of bio-Nano-Things, and deep-sea sightseeing, etc. A huge shift in technologies can also be seen among 4G, 5G, and 6G networks. 4G mainly used OFDM and MIMO; 5G uses mm-wave and massive MIMO, and 6G will use THz band and SM-MIMO, etc. to achieve its goal in terms of delay and throughput.

It is expected that in 6G networks, a fully ML and AI-enabled network will be in operation. That is why it is important to provide a detailed analysis of upcoming network architectures with their requirement in terms of AI.
Table \ref{table: AI} shows the link between various existing and future network architectures and related AI algorithms. SDN, NFV, and Cloud/fog/edge computing are considered potential network architectures that will use a fully AI-based environment to support 6G networks. Currently, these architectures partially support AI and ML aspects.

\begin{table*}[htbp]
  \caption{\label{tab: AIMech}Overview of AI mechanisms, their functional descriptions, and diverse applications where these AI mechanisms are transforming numerous industries, including healthcare, finance, and robotics. } 
    \label{table: new}
\begin{center}
\scalebox{0.85} {
  \begin{tabular}{p{4cm}|p{7.5cm}|p{7.5cm}}
\toprule
\textbf{AI Mechanism}                                         & \textbf{Description}                                                                      & \textbf{Applications}                                                     \\ \hline
Machine Learning \cite{9413429}                     & Algorithms that enable systems to learn from data and make decisions & Image recognition, fraud detection                  \\ \hline
Deep Learning \cite{hsu2021can}                     & Neural network models with multiple layers for complex tasks      & Speech recognition, autonomous driving             \\ \hline
Reinforcement Learning \cite{wurman2022outracing}              & Learning through interaction with an environment                     & Game playing (e.g., AlphaGo), robot control         \\ \hline
Natural Language Processing (NLP) \cite{liu2023pre}      & Processing and understanding human language                         & Language translation, chatbots                      \\ \hline
Computer Vision \cite{enguita2023transcriptomic}                    & AI for image and video analysis                                     & Object detection, facial recognition                 \\ \hline
Transfer Learning \cite{fabijanic2022automatic}                     & Applying knowledge from one domain to another                        & Pretrained models for different tasks               \\ \hline
Generative Adversarial Networks (GANs) \cite{guo2021towards} & Generating realistic synthetic data                                  & Image synthesis, deep fake generation                \\ \hline
Federated Learning  \cite{nguyen2022federated}                   & Training models on decentralized devices                            & Collaborative learning across edge devices           \\ \hline
Edge Computing \cite{jie202110}                      & Processing and analysis at the network edge                          & Real-time analytics, IoT data processing             \\ \hline
Quantum Computing  \cite{duong2022quantum}                    & Utilizing quantum phenomena for advanced computations                & Quantum encryption, optimization problems            \\ \hline
Swarm Intelligence \cite{khan2022swarm}                    & Collective behaviour of decentralized entities                       & Swarm robotics, swarm optimization                   \\ \hline
Explainable AI  \cite{fiandrino2022toward}                      & Making AI models transparent and interpretable                       & Visualizing model decisions, rule extraction         \\ \hline
Neuroevolution \cite{kong2022improving}                        & Evolving neural network architectures and parameters                 & Evolutionary robotics, neural architecture search    \\ \hline
Cognitive Computing \cite{rahmani2023cognitive}                   & Emulating human cognitive processes for decision-making             & Sentiment analysis, decision support systems         \\ \hline
Swarm Robotics \cite{cai2022self}                        & Coordinated behaviour of robot swarms                               & Multi-robot coordination, task allocation            \\ \hline
Knowledge Graphs \cite{zhou2022intelligence}                      & Structured representation of knowledge for reasoning and inference   & Semantic search, knowledge-based recommendations    \\ \hline
Probabilistic Graphical Models \cite{mao2023security}       & Modelling uncertainty and probabilistic relationships                & Bayesian networks, hidden Markov models             \\ \hline
Swarm Optimization  \cite{khan2022swarm}                   & Optimization inspired by collective behaviour                        & Particle swarm optimization, ant colony optimization \\ \hline
Robotic Process Automation \cite{adhikari20226g}            & Automating repetitive tasks using software robots                   & Data entry, invoice processing                       \\ \hline
Evolutionary Algorithms \cite{kouhalvandi2022overview}               & Optimizing solutions through genetic and evolutionary principles     & Genetic algorithms, evolutionary optimization        \\ \hline
Explainable Recommendation Systems  \cite{wu2022knowledge}   & Providing transparent and interpretable recommendations              & Explainable movie/music recommendations              \\ \hline
Self-organizing Networks  \cite{papidas2022self}             & Network organization and optimization based on decentralized control & Self-organizing network planning and management      \\ \hline
Genetic Programming  \cite{birabwa2022service}                  & Evolving computer programs to solve complex problems                 & Symbolic regression, program synthesis              \\ \hline
Knowledge-based Reasoning  \cite{kumari2022signature}            & Reasoning using explicit knowledge and rules                         & Expert systems, diagnosis and decision support       \\ \hline
Swarm Intelligence Optimization \cite{tong2022nine}       & Optimization inspired by swarm behaviour                            & Particle swarm optimization, ant colony optimization \\ \hline
Human-Machine Collaboration \cite{han2022multi}           & Collaboration between humans and AI systems                         & AI-powered assistants, collaborative robots         \\ \hline
Knowledge-based Systems \cite{zhou2022intelligence}               & Representing and using knowledge for decision-making                & Expert systems, rule-based reasoning                \\ \hline
Context-aware Computing  \cite{abdulqadder2022sliceblock}              & Utilizing contextual information for intelligent decision-making     & Context-aware recommendations, adaptive systems      \\ \hline
Bayesian Inference  \cite{wu2022knowledge}                  & Probabilistic inference using Bayes' theorem                       & Uncertainty quantification, Bayesian networks        \\ \hline
Metaheuristic Optimization  \cite{el2023metaheuristic}                                 & Iterative optimization algorithms inspired by natural processes                           & Genetic algorithms, simulated annealing                                   \\ \hline

\end{tabular}
}
\end{center}
\end{table*}

\section{6G Potential Use Cases} \label{sec:6}


The 6G network acts as a successor to the 5G cellular network. 6G networks will operate on higher frequencies than the 5G networks to provide higher bandwidth (in terabytes) and ultra-low latency (in microseconds). It is expected that 6G technology will bring a huge performance improvement for various existing communication technologies. For example, it is estimated that cellular data will expand 3-fold from 2016 to 2021 and there will be more than 125 billion connected cellular users globally by 2030 \cite{de2021survey}.

This section discusses the characteristics and requirements of some potential use cases of 6G networks. There are many research works that discuss the potential use cases for 6G technology \cite{xu20113d, zhang2018towards, pathak2015visible, lee2015cyber, wollschlaeger2017future, zanella2014internet, lu2014connected, choi2016millimeter, jafri2019wireless, denardis2020internet, snyder2017internet, show2021prospect, jiang2021road, al2021survey, ji2021survey, salh2021survey, dao2021survey, yaacoub2020key, wang20206g, alghamdi2020intelligent}. Based on the technology used, we categorize the potential use cases into three broader categories: VR, health, and Industry 5.0 as shown in Figure \ref{Figure:taxo}. These categories have various sub-categories which are discussed in this section.

\subsection{\bf Virtual Reality (VR)}
VR can be further divided into the following two categories:

{\bf Extended Reality (XR).} 
XR is a broader term that includes technologies to enhance our senses \cite{chuah2018and}. It mainly includes augmented reality (AR), mixed reality (MR), and virtual reality (VR). AR, MR, and VR over wireless links have been emerging as potential future applications targeting various areas such as education, training, entertainment, tourism, sport, and gaming.

AR takes a real-world environment and adds computer-based input to the environment. Hence, AR is a collaborative activity where real-world environment experience is obtained. The environment is built of enhanced objects and these objects are available in a real-world environment. AR technology makes use of various sensors to experience modalities like visual, haptic, olfactory, motion, and somatosensory. AR consists of three basic features: 3D visualization of objects, real-time collaboration, and a mixture of virtual and real-world environments \cite{ccoltekin2020extended}.

\begin{figure}[htb!]
        \centering
            \includegraphics[width=0.4\textwidth]{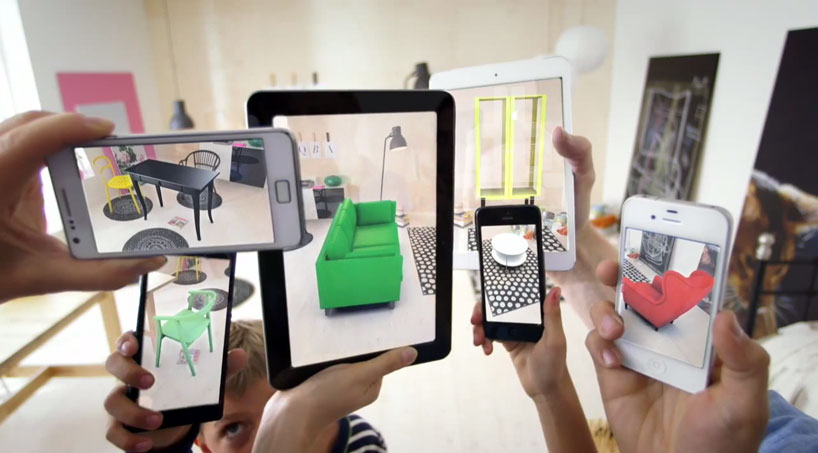}
            \caption{IKEA AR-based applications for customers \cite{carmigniani2011augmented}. }
            \label{Figure:AR}
\end{figure}

Figure \ref{Figure:AR} shows the AR-based user interface for the mobile applications of IKEA, where customers can try various options according to their real-world environment. For example, in a Dulux-based AR application, a customer can apply different colors in his real home environment, and in an IKEA application, a customer can try various pieces of furniture in a real home environment. Furthermore, these applications also handle the measurement and size issues where you need to deploy the furniture.

VR provides a purely virtual environment without considering the user's surroundings. VR-based devices such as HTC Vive and Google Cardboard provide users with a completely imaginary environment such as a squawking penguin colony with dragons. MR combines the characteristics of VR and AR. MR technology has started to take off with Microsoft’s HoloLens, one of the most famous mixed reality apparatuses \cite{wu2020integrated}.

These applications process exceptional QoS challenges including quality of experience, latency, and capacity/user. To achieve this, technologies like fog and cloud computing will help to act as computing and intelligence layers for AR/VR applications. 
4G has been proven as a key enabling technology for data-hungry applications like video-over-wireless. Later, the high use of these multimedia applications was managed by 5G using the mmW spectrum. The mmW spectrum attracted more data-intensive applications like AR, VR, and MR. It is expected that in the coming years, the 5G spectrum will be depleted due to the increasing use of these data-intensive applications as they demand system capacity up to 1 Tbps \cite{hindia2020platform}. One of the reasons for high system capacity is that the data used in AR/VR/MR-based applications cannot be compressed as it is presenting real-time interaction. Furthermore, managing real-time user interaction with AR/VR/MR-based data-intensive applications requires a delay in microseconds.

{\bf Holographic Telepresence.}   Holographic telepresence is a growing technology for fully interactive 3D video conferencing. Holographic telepresence systems are capable of presenting real-time, full-motion-based 3D images of people and objects in a room. It also accommodates real-time audio and associates it with a particular 3D image of a person. Images of people and surrounding objects at a remote location are initially captured, then compressed and transmitted over a broadband network. At the destination, the data is decompressed and similarly projected through laser beams as a conventional hologram is produced. Holographic telepresence is capable of revolutionizing various types of communication. For example, telepresence in teleoperation can allow medical specialists to engage in real-time complex operations from thousands of miles away. Furthermore, this technology can also reduce the travel requirement for personal and business meetings. Distance learning using telepresence can shape the future of universities and colleges. Other prominent examples include programmed movies, entertaining games, 3D navigation applications, etc. The applications mentioned above like a 3D holographic display require a set of high-level QoS requirements. According to a study, \cite{carmigniani2011augmented}, a raw uncompressed hologram having colors, and parallax, with 30 fps requires a data rate of up to 4.32 Tbps with a strictly bounded delay of a few milliseconds. 3D holographics require such a high data rate due to the requirement of thousands of synchronized angles. 6G network is capable of providing such high throughput and less delay required by these applications. Furthermore, an acceptable visual quality at the remote site is a challenging task.

\subsection{\bf Smart Healthcare} 
The primary healthcare challenges confronting the global population encompass a growing elderly demographic, escalating healthcare expenditures, and a substantial mortality rate stemming from chronic ailments \cite{akbar2018modelling, catarinucci2015iot}. Notably, in nations such as the United States, Germany, the United Kingdom, France, Switzerland, Japan, and Australia, life expectancy has risen to 79.3 years, 81 years, 81.2 years, 82.4 years, 83.4 years, 83.7 years, and 82.8 years, respectively. This upward trajectory in life expectancy is expected to exert significant pressure on existing healthcare systems and concomitantly elevate healthcare costs.

Furthermore, a substantial portion of the population dies from fatal and chronic diseases including hypertension, cardiovascular disease, diabetes, and asthma. Empirical investigations have unveiled that the impact of these maladies can be mitigated when detected in their nascent stages. To realize this objective, forthcoming healthcare systems must transform, leveraging digital and communication technologies to facilitate proactive wellness through early detection and continuous monitoring services.
It is estimated that the demand for remote healthcare monitoring services will burgeon to 761 million by the year 2025~\cite{kumar2020novel, zhu2019smart}. This underscores the imperative for advancing healthcare infrastructure and technology to address the burgeoning healthcare needs of an aging and chronically ill population while containing costs. 

Smart healthcare services will encompass a broad spectrum of functionalities, ranging from continuous monitoring to advanced telemedicine solutions, such as real-time remote surgical procedures. While the current 5G spectrum has initiated its role in enhancing remote healthcare capabilities, certain demanding scenarios, such as remote surgeries, comprehensive health monitoring within elderly care facilities equipped with an array of extensive biomedical sensors, and the provision of remote healthcare services to numerous soldiers on battlefields, necessitate ubiquitous health monitoring characterized by ultra-high data rates measured in gigabits per second (Gbps) and terabits per second (Tbps), as well as ultra-low latency to support real-time solutions. In this context, 6G networks stand poised to revolutionize remote healthcare by surmounting the limitations of time and space\cite{zhang2018towards}.

The technological advancements inherent to 6G networks will usher in a new era for eHealth applications, harnessing groundbreaking technologies such as mobile edge computing, artificial intelligence (AI), fog computing, and cloud computing.

\begin{figure}[htb!]
        \centering
            \includegraphics[width=0.5\textwidth]{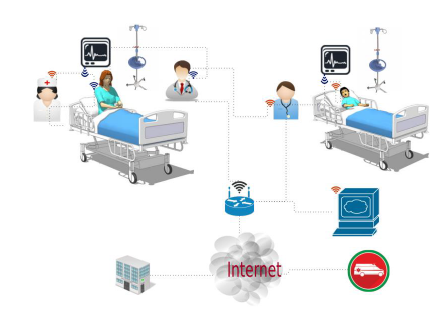}
            \caption{Remote patient monitoring in a hospital}
            \label{Figure:SmartH}
\end{figure}

Figure \ref{Figure:SmartH} shows a typical remote patient monitoring system for hospitals where patients, nurses, and medical devices are equipped with various healthcare sensors, and the monitoring data is sent to the Internet and cloud computing architectures for further processing. The doctors can see the different results online and can monitor and diagnose the patients in real-time from any remote location.

{\bf Brain–Computer Interaction.} A brain-computer interface (BCI) is a communication pathway between a sensor-equipped and an external device. The main objectives of BCIs include repairing human cognitive or sensory-motor functions \cite{tripathy2016application, jafri2019wireless}. The sensors collect the brain signals and send them to a digital device for analysis.

Research on BCIs started at the University of California, Los Angeles (UCLA) in the early 1970s,  under research funding from the National Science Foundation. Recent studies in the area of  Human-Computer Interaction (HCI) show that machine learning algorithms can play a key role in successfully extracting features from the frontal lobe and then classifying mental states (Normal, Neutral, focused) and emotional states (Good/bad mood) using EEG generated brain-data \cite{tripathy2016application, jafri2019wireless}. 
Such communication requires ultra-reliable communication with very low delay, values in microseconds.  The characteristics of 6G architecture will support the deployment of BCI systems.

{\bf Human Senses-Based Applications.} 
Humans have five basic senses: hearing, vision, taste, smell, and touch. In the future, applications will be available that will be capable of transferring human senses data to a remote location for processing. Such applications use sensor-based technologies which make use of neurological methods. It requires ultra-low delay and high throughput. The 6G networks can provide such services \cite{de2021survey, shahraki2021comprehensive}. 

\subsection{\bf Industry 5.0 } 

Industry 5.0 or the Fourth Industrial Revolution, is the recent revolution of traditional manufacturing using smart technology. Technically, it mainly emphasizes the usage of large-scale machine-to-machine communication (M2M) with the help of Internet of Things (IoT) architectures and deployments to obtain automation \cite{show2021prospect}. Industry 5.0 will make the machine's components capable of performing self-diagnostic and predicting any possible future issues using AI algorithms so that any future failure can be avoided. Such decisions constitute the first steps of the Industry 5.0 era, where machines will make autonomous decisions and these decisions will have significant effects on people's lives. High achievements obtained in many areas with AI algorithms are the basis of these developments in the industry.

\begin{figure}[htb!]
        \centering
            \includegraphics[width=0.45\textwidth]{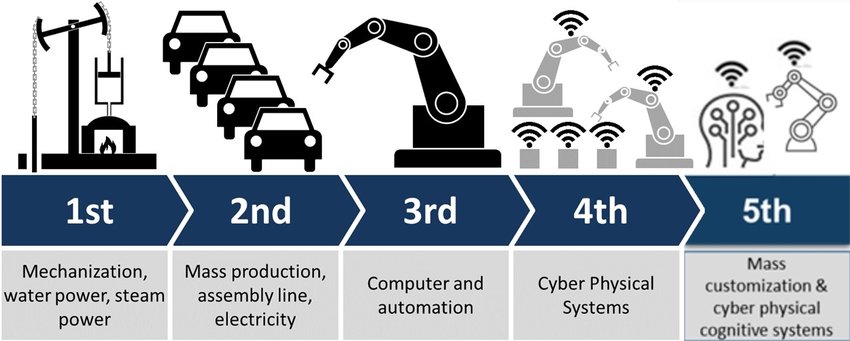}
            \caption{Automation under Industry 5.0 \cite{broadband_img3}. }
            \label{Figure:Idustry55}
\end{figure}

Furthermore, machine components should be capable of communicating with the other components and other machines. Predictive maintenance, IoTs, smart sensors, and 3D printers are the main key enablers for Industry 5.0. 6G technology will further promote the Industry 5.0 concept which was initiated with 5G. Industry 5.0 introduced the idea of the digital transformation of manufacturing through cyber-physical systems (CPS), which opens the gates between physical factories and virtual computation. Such a combination will promote new emerging technologies like M2M communications in a cost-effective way \cite{ show2021prospect}. Efficient automation systems in terms of asynchronous communication are considered key enablers of Industry 5.0, in this regard, 6G is proposing the use of terahertz domains to provide the required QoS for smart integrated systems. Figure \ref{Figure:Idustry55} shows the evolution of Industry 5.0. 

{\bf Indoor Coverage.} 
The 5G infrastructure usually operates under the milimeterWave spectrum (60 GHz) for indoor scenarios; however, it faces issues like penetration in solid materials. 5G  also proposed the use of distributed antenna systems (DASs) for indoor environments which poses issues like scalability and deployment. 6G focuses on cost-efficient solutions for indoor communication scenarios using ultra-high capacity wireless relays with the visible light spectrum \cite{pathak2015visible}.

{\bf Smart City.} 
A smart city can be defined as an urban area equipped with various types of sensors that collect data and then perform analysis of the collected data to manage and improve the infrastructure, services, and quality of life. The collected data may belong to various resources including citizens, installed devices, and buildings. The embedded data analytical capabilities help to manage traffic systems, power plants, schools, libraries, hospitals, water supply networks, waste handling, crime detection, etc \cite{kim2017smart}.

5G technologies act as a key enabler for a smart city; however, 5G only made the city partially smart which means that inside a city, only the fragments (traffic management, water management, electricity, etc.)
are individually smart but not holistically as a smart city. 6G technologies have the potential to speed up the adoption of smart cities by offering required QoS using innovative and disruptive communication technologies \cite{zanella2014internet}.

 6G will provide support for user-centric M2M transmissions to further improve the deployment of smart cities. Furthermore, to provide a long battery life for the devices in smart cities, energy harvesting approaches will be used, though, 5G initially promised to adopt energy harvesting solutions but still, it is not being adopted. 

{\bf Internet of Vehicular Services.} 
The creation of a highly networked and intelligent transportation system is a compelling use case for the Internet of Vehicles (IoV) utilizing 6G networks. This system uses distributed intelligence and advanced communication technologies to improve traffic efficiency, road safety, and the entire driving experience.
Unmanned vehicular mobility refers to vehicular mobility without a driver/pilot on board. These vehicles can be controlled remotely and they are capable of operating and navigating autonomously by sensing their environment. There are different types of unmanned vehicles including Unmanned ground vehicles (UGV) like cars and combat vehicles, Unmanned aerial vehicles (UAV) such as drones and autonomous spaceport drone ships, and unmanned underwater vehicles (UUV) and driver-less trains. 

\begin{figure}[htb!]
        \centering
            \includegraphics[width=0.45\textwidth]{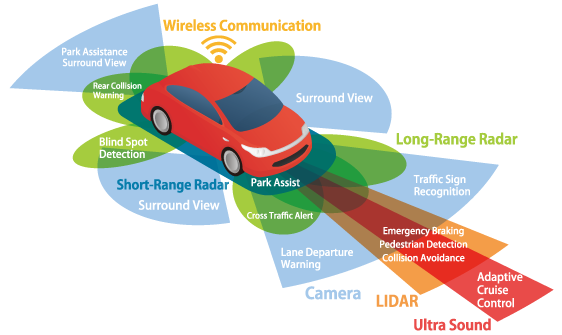}
            \caption{An unmanned car with numerous sensors \cite{khan2021role}.}
            \label{Figure:Selfcar}
\end{figure}

The locomotive industry is precipitously advancing towards fully autonomous transport systems which will ease the management burden on the organization \cite{lu2014connected}. Connected and autonomous vehicles (CAVs) are one of the potential use cases for future unmanned mobility. However, its deployment is still challenging in terms of time-bounded services and high mobility (up to 1000 Km/h). Further, it is estimated that vehicles will be fitted out with a high number of sensors (greater than 200 per vehicle by 2020) which can be only served with terahertz spectrum \cite{choi2016millimeter}. Figure \ref{Figure:Selfcar} shows a self-driving car example in Australia with all embedded features and requirements. Moreover, flying vehicles like drones indicate a massive potential for several use cases: industrial construction, the agriculture sector, and entertainment. 6G can provide services for such massive data-hungry use cases.

{\bf Internet of Everything.}  
The term IoE evolved a few years ago, and there is a clear difference between the definition of IoE and IoT. IoE is defined as intelligent connectivity among things, data, processes, and people. The data is converted into useful information and delivered to the right person/object at the right time. Billions of objects update their status by using various sensors. The IoT can be defined as the network of physical objects that are accessed on the Internet. 
IoE incorporates a massive network of things where each thing-network is an autonomous system and capable of coordination with other networks \cite{denardis2020internet}. IoT and IoE have the same purpose in terms of connectivity; however, IoE’s terminology is used for a broader aspect with massive data. IoE also introduces the intelligence layer for decision-making. 
The IoE is expected to provide massive connectivity in various domains including healthcare, agriculture, logistics, manufacturing, and education. The 6G networks will support such massive data connectivity using its flexible heterogeneous architecture \cite{snyder2017internet}.
\subsection{\bf Communication networking technologies: Underwater and Terrestrial Networks }

{\bf Underwater Networks.} Underwater networks are seeing new opportunities as a result of the advancement of 6G technology. A noteworthy application of 6G networks to improve the effectiveness and dependability of data transfer in marine environments is the integration of underwater communication devices. Underwater exploration, environmental monitoring, and maritime security are just a few of the operations that this integration can greatly enhance. To tackle issues like scarce wireless resources and high connection demands in underwater locations, for instance, 6G can be combined with multiple-input multiple-output (MIMO) systems and non-orthogonal multiple access (NOMA) technology \cite{vinogradov2016key}.
Drones for underwater communication is another exciting application. In order to link underwater sensor networks with terrestrial base stations and enable real-time data transmission and monitoring, these drones can serve as relay nodes. Applications needing low latency and high reliability, such environmental monitoring and search and rescue, may find this approach especially helpful.

{\bf Terrestrial Network Technologies.} 

6G technology promises ultra-low latency, huge connection, and hitherto unheard-of data rates, which might completely transform terrestrial networks. A significant use for 6G is in the field of smart cities, where it can facilitate the widespread installation of IoT devices and provide cutting-edge features like real-time surveillance, smart grid management, and driverless cars. Reconfigurable intelligent surfaces (RIS) and unmanned aerial vehicles (UAVs) are two examples of technologies that can be integrated with 6G to significantly improve network coverage and capacity, guaranteeing smooth connectivity in metropolitan areas.
Disaster management and emergency response represent yet another important use. In places where the current infrastructure has been destroyed or is nonexistent, 6G-enabled UAVs can swiftly set up communication networks. The coordination and effectiveness of relief efforts can be enhanced by the ad hoc networks that these UAVs can create, which will allow rescue teams and control centres to communicate in real time \cite{zanella2014internet}.

\subsection{\bf Summary}

Section V provides a categorization of potential use cases for 6G networks, the categorization includes VR, health, and Industry 5.0. VR is further subdivided into two categories including XR and teleportation. Both XR and teleportation are discussed in detail with examples. The second important category belongs to healthcare and a couple of potential use cases are discussed under this category including BCI and human sense-based applications. A discussion about the implementation of BCI for the 6G network is provided. Industry 5.0 is discussed as the last category of the potential use cases that maps AI and ML with industrial applications. Industry 5.0 is being considered a revolution for the automated industry and it is vital to link it with upcoming 6G networks.

\section{Future Research Directions}\label{sec:8}

Numerous technical issues need to be resolved for the successful deployment of 6G networks. Key concerns include the development of cutting-edge AI and machine learning algorithms for real-time network optimization and predictive maintenance, the integration of holographic and haptic communication technologies for immersive telepresence, and the exploration of the THz spectrum for ultra-high-speed data transmission. Furthermore, it is essential to conduct research on quantum communications, particularly quantum key distribution (QKD) for enhanced security, and to create sustainable, energy-efficient network protocols. Emphasizing strong security and privacy measures in post-quantum cryptography contexts, as well as achieving global connectivity through the integration of low earth orbit (LEO) satellites, will also be crucial. These directions highlight a multidisciplinary approach while addressing the necessary technological advancements. We argue the following research areas should be focused for future research:

{\bf High-Level Propagation and Atmospheric Absorption of THz}. The THz frequency bands offer very high data rates. However, it faces a significant challenge in data transmission over long distances due to high propagation loss caused by atmospheric absorption characteristics. These absorptive effects are influenced by the often variable and unpredictable atmospheric conditions\cite{markit2017internet}. To address this issue, a new transceiver architecture with efficient antennas for the THz frequency bands is required. Additionally, health and safety concerns related to the THz band need special attention from both academia and industry. 

{\bf Spectrum and Interference Management.} The shortage of spectrum at higher frequencies and the associated interference problems necessitate efficient and innovative spectrum-sharing schemes. Effective spectrum management is crucial for achieving optimal resource utilization and maximizing quality of service (QoS). To ensure the efficient deployment of 6G, researchers must address issues such as spectrum management for heterogeneous networks and devices operating within the same frequency band~\cite{matinmikko2020spectrum}. 

{\bf Beam Management in THz Communications.} Beamforming via advanced MIMO systems is a potential technology to support high data rates. However, managing beams at high-frequency bands like the THz band is challenging due to the unique propagation characteristics. Therefore, designing an effective beam management scheme to handle these propagation characteristics in advanced MIMO systems is crucial. Furthermore, for seamless handover in high-speed vehicular systems, it is important to efficiently select the optimal beam \cite{markit2017internet}.

{\bf Device Capability.} The initiation of the THz spectrum will introduce new sensing capabilities to 6G devices, including high-definition imaging and frequency spectroscopy. These enhanced sensing capabilities, when combined with high precision, will play a crucial role in understanding the context and could help build a trusted and secure environment. The 6G system will require more advanced capabilities from devices to operate efficiently. For example, advanced smartphones equipped with AI, XR and integrated sensing must be resourceful enough to enable the new features of 6G, such as a throughput of 1 Tb/s. Existing 5G devices may have limitations in supporting some of the 6G features. Therefore, a large number of devices currently connected to 5G will need to be compatible with 6G.

{\bf High Capacity Backhaul Network Connectivity.} 
One significant element in the connectivity solution for 6G networks is an efficient backhaul connection. Designing an appropriate backhaul for remote areas can be very challenging. The high-density access networks envisioned for 6G are distinct and pervasive, capable of providing ultra-high data rates. In the future, the backhaul network will play a critical role in providing massive connectivity between core and access networks. Evolving satellite constellations, such as Starlink by SpaceX, and systems by OneWeb and Telesat, have significant potential to support such scenarios; however, they are expensive solutions. Optical fiber and free-space optical (FSO) networks are more appropriate options and have the potential to further enhance the performance of backhaul networks in 6G. Therefore, advancements in backhaul networks are essential~\cite{sharma2021review}.

{\bf Complexity in Resource Management for 3D Networking.} 3D networking enables the creation of stunning 3D network diagrams for a better understanding of network structures. It is considered one of the powerful future aspects of networking that will merge with 6G networks. 3D-based networks expand in the vertical direction, necessitating innovative methods for resource management, routing protocols, mobility support, and channel access. Significant attention from both industry and academia is required to implement this concept. Merging with 6G networks, 3D networking will enhance networking operations and management \cite{huq20213d}.

{\bf Heterogeneous Hardware Constraints.} 
The 6G network is envisioned as a massive collection of heterogeneous devices, frequency bands, architectures, typologies, and operating systems. Additionally, hardware operational settings for mobile terminals and signaling devices will differ substantially. The MIMO technique will be more advanced in 6G, featuring a more complex architecture than in 5G. Moreover, 6G will need to adapt to intricate communication protocols with sophisticated algorithms. The integration of advanced AI and machine learning techniques, such as unsupervised and reinforcement learning, could further complicate hardware operations. Consequently, incorporating all these heterogeneous aspects into a single platform will be  challenging~\cite{tan2020thz}. 

{\bf Autonomous Wireless Systems.}  
Designing autonomous wireless systems with simultaneous responses in real time is a complex task that cannot be achieved through incremental changes to existing control and optimization methodologies. It requires a fundamental shift by integrating machine intelligence into the wireless architecture and infrastructure. Autonomous wireless systems in 6G networks will encompass several heterogeneous technological components, including computing, machine learning, autonomous cloud computing systems, and diverse wireless systems~\cite{elliott2019recent}. Each of these components necessitates specific quality of service parameters to create a fully autonomous system, such as ML capabilities, rapid data processing, extensive storage capacity, high internet availability, and robust security services. With the advanced infrastructure and features of 6G networks, it is anticipated that fully autonomous systems will be realized for various applications, including cars, UAVs, drones, and more.

UAVs are emerging as a potential 6G application due to their ability to provide ultra-high data rates and ubiquitous wireless connectivity. UAVs offer special features such as ease of deployment, high line-of-sight link probability, and significant degrees of freedom enabled by their controlled mobility. They can be deployed to extend wireless connectivity in disaster scenarios, such as natural disasters. One of the most valuable features of UAVs is their ability to adaptively reposition according to environmental changes and user demands. The objective of connecting the independent components through 6G will be facilitated by aerial base stations due to their simple deployment.

\section{Conclusion} \label{sec:9}

In this survey, we initially evaluated the key successes and challenges of 1G to 5G and then focused on the potential applications, enabling technologies, specifications, and requirements, and AI-based wireless applications and technologies of 6G. The 5G communication system is expected to be fully deployed globally in 2022. However, the current 5G technologies are not capable of catering to the increasing demand for the future wireless communication which is envisioned for 2030. 

Research activities on 6G technologies and systems are 
still 
in their early stages. This survey envisions the options and methods to accomplish the objectives of 6G wireless communication. In this paper, we also presented the potential applications for 6G with key enabling technologies so that stringent QoS requirements of future applications can be fulfilled. These requirements are ultra-high throughput (1Tb/s to 20 Tb/s),  extreme-low latency (tens of microseconds), high reliability, ubiquitous network connectivity, and very low power consumption. To support such QoS, advanced network architectures and technologies like fog computing, cloud computing, software-defined networks, optical wireless networks, network virtualization, and THz spectrum communications are required. Moreover, future applications will generate massive data, and to handle such big data, advanced AI techniques and algorithms are required. We summarized some of the potential AI methods for specific network architectures. Along with explaining the vision and goal of 6G communications, we have identified the numerous technologies that could be used for 6G communication. We also discussed the probable challenges and research directions to achieve the goals for 6G.

\balance

\bibliographystyle{elsarticle-num-names}
\bibliography{main}

\end{document}